\documentclass[12pt,aps,tightenlines,epsfig,graphicx,rotate,floats,
nofootinbib]{revtex4}

\usepackage{epsfig}%
\usepackage{graphicx}
\usepackage{dcolumn}
\usepackage{bm}

\renewcommand{\d}{\delta}

\newcommand{\bea}{\begin{eqnarray}}
\newcommand{\eea}{\end{eqnarray}}
\newcommand{\beq}{\begin{equation}}
\newcommand{\eeq}{\end{equation}}

\newcommand{\gsim}{~{}_{\textstyle\sim}^{\textstyle >}~}
\newcommand{\lsim}{~{}_{\textstyle\sim}^{\textstyle <}~}
\newcommand{\ba}{\begin{array}{c}}
\newcommand{\bat}{\begin{array}{cc}}
\newcommand{\ea}{\end{array}}

\def\slashchar#1{\setbox0=\hbox{$#1$}\dimen0=\wd0%
\setbox1=\hbox{/}\dimen1=\wd1%
\ifdim\dimen0>\dimen1%
\rlap{\hbox to
\dimen0{\hfil/\hfil}}#1\else                                     
\rlap{\hbox to \dimen1{\hfil$#1$\hfil}}/\fi}
\newcommand{\beqa}{\begin{eqnarray}}
\newcommand{\eeqa}{\end{eqnarray}}
\begin{document}



\title{
Lepton Flavor Violation without Supersymmetry
}
\author{V. Cirigliano} 
\author{A. Kurylov } %
\author{M.J. Ramsey-Musolf}%
\author{P. Vogel}%
\affiliation{
Kellogg Radiation Laboratory, California Institute of Technology, 
Pasadena, CA 91125, USA 
}

\date{\today}

\begin{abstract}

We study the lepton flavor violating (LFV) processes $\mu\to e\gamma$,
$\mu\to 3e$, and $\mu\to e$ conversion in nuclei in the left-right
symmetric model without supersymmetry and perform the first complete
computation of the LFV branching ratios $B(\mu \to f)$ to leading
non-trivial order in the ratio of left- and right-handed symmetry
breaking scales. To this order, $B(\mu\to e\gamma)$ and $B(\mu\to e)$
are governed by the same combination of LFV violating couplings, and
their ratio is naturally of order unity.  We also find $B(\mu\to 3
e)/B(\mu\to e) \sim 100$ under slightly stronger assumptions. 
Existing limits on the branching ratios already substantially 
constrain mass splittings and/or mixings in the heavy neutrino sector.  
When combined with future collider studies and precision electroweak 
measurements, improved limits on LFV processes will test the 
viability of low-scale, non-supersymmetric LFV scenarios.

\end{abstract}

\pacs{Valid PACS appear here}
\maketitle


\section{Introduction}

Leptons of different flavors do not mix in the Standard Model (SM) of
electroweak interactions as a consequence of vanishing neutrino
masses. The observation of neutrino oscillations, however, has
provided clear evidence that nature does not conserve lepton flavor
and that the SM must be part of a more fundamental theory that allows
for lepton flavor violation (LFV). In the most widely held theories of
neutrino mass, LFV is generated at high scales that are well beyond
the reach of present and future collider experiments. Searching for
LFV among charged leptons in low-energy measurements is an alternative
and important way to probe additional aspects of LFV at such high
scales.

Attempts to observe, and theoretically predict, the manifestations of
LFV involving various modes of muon decay have a long tradition. The
rather small upper limit ($2 \times 10^{-5}$) on the branching ratio
for the $\mu \rightarrow e + \gamma$ decay determined by Lokanathan
and Steinberger \cite{Steinberger55} almost fifty years ago led to a
flurry of theoretical activity (see e.g \cite{Feinberg58} and many
subsequent papers) that resulted in the realization that the electron
and muon neutrinos are different particles -- a fact confirmed
experimentally shortly afterward.  Over the intervening years the
increase in the intensity of muon beams and advances of experimental
techniques led to impressive improvement in the sensitivity of various
searches for LFV. Even though no positive effects have been seen so
far, the upper limits of the corresponding branching ratios became
smaller by a factor of $\sim 10^6$. At present, the most stringent
limit on the branching ratio for $\mu\to e\gamma$ is~\cite{muegamma99}
\beq
\label{eq:MEGA}
B_{\mu\to e\gamma}\equiv {\Gamma(\mu^+\to e^+\gamma)\over\Gamma(\mu^+\to e^+ 
\nu{\bar\nu})} < 1.2\times 10^{-11}\ \ \ {\rm 90\% C.L.}\ \ \ ,
\eeq
obtained by the MEGA collaboration, while for the process of $\mu\to
e$ conversion in gold nuclei, the SINDRUM collaboration has obtained
the limit~\cite{mueconvAu}
\beq
\label{eq:SINDRUM}
B_{\mu\to e}^{A}\equiv {\Gamma(\mu^- + A(N,Z) \to e^- + A(N,Z))\over 
\Gamma(\mu^- +A(Z,N)\to \nu_\mu+ A(Z-1, N+1))} < 
8 \times 10^{-13}\ \ \  {\rm 90\% C.L.}\ \ \  .
\eeq
The present limits on other branching ratios are similarly impressive:
$1.0 \times 10^{-12}$ for $B_{\mu^+ \rightarrow e^+e^-e^+}$
\cite{mu3e88}, $4.3 \times 10^{-12}$ for $B_{\mu\to e}^{\rm
Ti}$\cite{mueconvTi}, and $4.6 \times 10^{-11}$ for $B_{\mu\to e}^{\rm
Pb}$\cite{mueconvPb}.  Two ambitious new experiments aiming at
substantial improvement in the sensitivity are being developed: MEG
plans to reach sensitivity of $\sim 5 \times 10^{-14}$ for $B_{\mu^+
\rightarrow e^+ + \gamma}$ \cite{MEG}, while MECO aims to reach $\sim 5
\times 10^{-17}$ for $B_{\mu\to e}^{\rm Al}$ in aluminum \cite{MECO}.

Theoretically, the focus in recent years has been on frameworks that
could both account for neutrino mass generation at high scales and
lead to observable LFV in future experiments with charged leptons.
The direct effects of light neutrinos on charged lepton LFV are \lq\lq
GIM suppressed" by a factors of $(\Delta m_\nu^2/M_W^2)^2\lsim
10^{-50}$ in the rate and are, thus, entirely negligible. In order to
obtain LFV effects that could be seen by experiment, a mechanism must
exist for overcoming this GIM suppression. Such a mechanism
necessarily involves physics at mass scales heavier than the weak
scale. The primary motivation for LFV studies involving charged
leptons is to help determine both the relevant scale as well as the
most viable models associated with it. 

Although a variety of such models have been considered, based on
various supersymmetry (SUSY)
scenarios~\cite{Borzumati:1986qx,Leontaris:1985pq,Hisano:1995nq,
Barbieri:1994pv,Barbieri:1995tw,Huitu:1997bi}, or left-right
symmetry~\cite{Riazuddin:hz,Barenboim:1996vu}, 
the most commonly-quoted are 
SUSY grand unified theories
(GUTs), wherein quarks and leptons are assigned to the same
representation of the unification gauge group at the GUT
scale. Consequently, the large Yukawa coupling responsible for the top
quark mass also appears in LFV
couplings\cite{Barbieri:1994pv,Barbieri:1995tw}. The latter then give
rise -- via renormalization group evolution -- to sizable lepton
flavor non-diagonal soft SUSY-breaking terms at the TeV
scale. Superpartner loops that contain insertion of these terms then
produce unsuppressed LFV transitions involving charged leptons.  For
example, in a SUSY SU(5) scenario, one has~\cite{Barbieri:1994pv}
\bea
B_{\mu\to e\gamma}& = & 2.4\times 10^{-12}
\left(\frac{|V_{ts}|}{0.04}\frac{|V_{td}|}{0.01}\right)^2\left(\frac{100\
{\rm GeV}}{m_{\tilde\mu}}\right)^4\\ B_{\mu\to e}^{\rm Ti} & = &
5.8\times 10^{-12} \ \alpha\
\left(\frac{|V_{ts}|}{0.04}\frac{|V_{td}|}{0.01}\right)^2\left(\frac{100\
{\rm GeV}}{m_{\tilde\mu}}\right)^4\ \ \ ,
\eea
neglecting gaugino masses. For superpartner masses of order the weak
scale, one would expect to see non-zero signals in the up-coming MECO
and MEG experiments under this scenario. Moreover, one would also
expect to observe an order $\alpha$ suppression of $B_{\mu\to e}^{A}$
relative to $B_{\mu\to e\gamma}$ in this case since the conversion
process entails the exchange of a virtual gauge boson between leptons
and the nucleus rather than emission of a real photon.

In this paper, we study an alternative paradigm for LFV, wherein LFV
occurs at much lower scales and does not require the presence of
supersymmetric interactions to overcome the GIM suppression factor. In
this scenario, neutrino mass generation occurs at the multi-TeV
scale via a spontaneously broken extended gauge group\footnote{For a
discussion of neutrino masses in this scenario and related
phenomenological issues, see, {\em e.g.},
Refs. \cite{Deshpande:1990ip,Mohapatra:2003qw}.}.  LFV for charged
leptons arises from the interactions of the additional gauge bosons,
heavy neutrinos, and Higgs bosons associated with the extended gauge
symmetry.  As an explicit realization of this scenario, we work within
the left-right symmetric model
(LRSM)~\cite{LRSM1,Mohapatra:1979ia,Mohapatra:1980yp}, which gives a
minimal, non-supersymmetric extension of the SM with non-sterile,
right-handed Majorana neutrinos.  As such, it contains triplet Higgs
fields that have non-zero hypercharge and that provide the simplest
mechanism for generating a Majorana mass term~\cite{Gelmini:1980re}.
As pointed out in Ref. \cite{Raidal:1997hq}, models of this type may
give rise to unsuppressed operators for the LFV decay $\mu\to 3e$, and
these operators in turn induce logarithmically enhanced amplitudes for
$\mu\to e$ conversion at loop 
level\footnote{Specific realizations of these ideas
been discussed for a doubly-charged scalar singlet~\cite{Raidal:1997hq}
and R parity-violating SUSY~\cite{Huitu:1997bi}.}. In the present case
such effects -- which were missed in earlier LRSM studies
\cite{Riazuddin:hz,Barenboim:1996vu} -- result from the presence of
the triplet Higgs fields.  The large logarithms can compensate for the
${\cal O}(\alpha)$ suppression of $B_{\mu\to e}^{A}$ relative to
$B_{\mu\to e\gamma}$ that generically follows for SUSY GUTs, and for
Higgs masses of order 10 TeV or below, both branching ratios may be
large enough to be seen in future measurements. Roughly speaking, we
find
\bea
B_{\mu\to e\gamma}& \approx & 10^{-7}\times |g_{\rm lfv}|^2 \left({1\
{\rm TeV}\over M_{W_R}}\right)^4\\ B_{\mu\to e}^{\rm Al} & \approx &
10^{-7}\times \alpha |g_{\rm lfv}|^2 \left({1\ {\rm TeV}\over
M_{\delta_R^{++}}}\right)^4 \left(\log{M_{\delta_R^{++}}^2\over
m_\mu^2}\right)^2
\eea
where $M_{W_R}$ is the mass of the right-handed charged gauge boson,
$M_{\delta_R^{++}}$ is the mass of a doubly charged, SU(2)$_R$ triplet
Higgs, and
\beq 
g_{\rm lfv} = \sum_N
 \Bigl(K_R^\dag\Bigr)_{eN}\Bigl(K_R\Bigr)_{N\mu}\left(M_N/M_{W_R}\right)^2\
 \ , 
\eeq
with $K_R$ being a flavor mixing matrix for the right-handed neutrinos
of masses $M_N$ (see below). In the limit of degenerate, right handed
neutrinos, the LFV factor $g_{\rm lfv}=0$. For heavy masses at the 
TeV scale, present experimental limits already  
constrain this factor to be tiny: 
$|g_{\rm lfv}|\lsim 10^{-2}$. In this case, the heavy neutrino
spectrum must either be nearly degenerate or devoid of significant
flavor mixing.

Note that both branching ratios are proportional to the same LFV
factor, $|g_{\rm lfv}|^2$. Naively, one would expect the loop graphs
giving rise to $\mu\to e\gamma$  (with heavy neutrino-gauge
boson intermediate states) and the logarithmically enhanced loops that
dominate $\mu\to e$ to have different prefactors. As discussed in more
detail below, however, the doubly charged, triplet Higgs
$\delta_R^{++}$ and its left-handed companion can have LFV violating
Yukawa couplings $h_{ij}$ of ${\cal O}(1)$, but the sum over
intermediate states in the logarithmically enhanced loop graphs
converts the sum over products of these couplings into $g_{\rm lfv}$.

Should $g_{\rm lfv}$ turn out to be nonzero, then one would expect the
two branching ratios to be of similar size since the product of the
$\ln^2$ and $\alpha$ in $B_{\mu\to e}$ is ${\cal O}(1)$. We expect
that any theory with non-sterile heavy Majorana neutrinos will contain such log
enhancements, due to the presence of a more complicated Higgs sector
than one finds in the SM. However, in SUSY GUT scenarios where LFV
occurs at high scales, these logarithmically enhanced loop effects
decouple below the GUT scale and do not affect the relative magnitudes
of the branching ratios. Only when the symmetry-breaking scale is
relatively light does one expect the two branching ratios to be
commensurate in magnitude.

Somewhat weaker statements about the relationship between $B_{\mu\to
e}$ and $B_{\mu\to 3e}$ can also be made within the context of this
model. In particular, we find
\beq B_{\mu\to 3e} \approx 300 \times {|h_{\mu e}h_{ee}^*|^2 \over
|g_{\rm lfv}|^2} \ \times B_{\mu\to e}^{\rm Al}\ \ \ , \eeq
so that if all of the triplet Higgs couplings $h_{ij}$ are of roughly
the same size and no cancellations occur in the sum 
$g_{\rm lfv}= \sum_j h_{\mu j}
h_{j e}^*$, the $\mu\to 3e$ branching ratio should be roughly two
orders of magnitude larger than the conversion ratio. Given the
present experimental limits on $B_{\mu\to 3e}$, one would then expect
$B_{\mu\to e}^{\rm Al}$ to be of order $10^{-14}$ or smaller. As we
discuss below, if the conversion ratio is found to be non-zero with
significantly larger magnitude, then one would also expect to see a
sizable effect in the channel $\tau\to 3\ell$ (where $\ell$ 
denotes a charged lepton). 

Finally, we observe that, while the logarithmic enhancement of
$B_{\mu\to e}^{A}$ is a generic feature of any model that yields
effective $\mu\to 3e$ operators at tree level, precise relationships
between the various LFV observables depend on details of the model. In
this respect, our perspective differs somewhat from the view in
Ref. \cite{Raidal:1997hq}. Indeed, the presence of a common factor of
$|g_{\rm lfv}|^2$ in $B_{\mu\to e\gamma}$ and $B_{\mu\to e}^{A}$ --
but not $B_{\mu\to 3e}$ -- and its relation to the heavy neutrino
spectrum follows from the pattern of symmetry breaking in
this scenario and the corresponding hierarchy of scales that enters
the couplings of the right-handed gauge sector to matter. In order to
implement this hierarchy in a self-consistent way, we adopt a power
counting in $\kappa/v_R$, where $v_R$ and $\kappa$ are the scales,
respectively, at which SU(2)$_R$ and electroweak symmetry are
broken. In contrast to previous
studies~\cite{Riazuddin:hz,Barenboim:1996vu}, we compute all LFV
contributions through leading, non-trivial order in $\kappa/v_R$ and
show that they decouple in the $v_R\to\infty$ limit as one would
expect on general grounds~\cite{Appelquist:tg}. In addition, we point
out the prospective implications of other precision measurements and
future collider studies for LFV in this scenario and vice-versa. The
identification of such implications necessarily requires the adoption
of a specific model, as the corresponding symmetries of the model
dictate relationships between the coefficients of effective operators
that would appear in an effective field theory framework. Thus, it is
useful to have in hand a comprehensive treatment within various model
frameworks in order to use experiment to discriminate among them.  In
R parity-violating SUSY, for example, the LFV couplings that generate
$\mu\to e$, {\em etc.} also appear, in general, in the mass matrices
for light neutrino flavors~\cite{grossman:2003}, whereas in the LRSM
LFV for charged leptons and light neutrinos are effectively
independent.

Our discussion of the calculation is organized in the remainder of the
paper as follows. In Section II we review the main features of the
LRSM and define the relevant quantities.  In Section III the
effective vertices are calculated and the effective Lagrangians
for the LFV processes are determined. Some of the detailed formulae
are collected in the Appendices. Section IV gives an analysis of the
results, along with a discussion of the rates as well as their ratios. We
conclude in Section V.

\section{The model}    

The gauge group of the theory is $\rm SU(2)_L\times SU(2)_R\times
U(1)_{B-L}$ with the gauge couplings $g_L=g_R=g$ for the two ${\rm
SU(2)}$s and $g^\prime$ for the ${\rm U(1)}$.  In this paper we
follow the notation developed in Ref.~\cite{Duka:1999uc} where the   
LRSM, its quantization, and its Feynman 
rules are discussed in detail. 
Below, we give a very brief introduction to the model, and explicitly
define the quantities used in subsequent analysis. 

The matter fields of the model include leptons (L$_{L,R}$) and
quarks (Q$_{L,R}$), which are placed in the following
multiplets of the gauge group:
\beqa
\label{eq:matter}
L_{iL}&=&\left( \begin{array}{c}
\nu^\prime_i\\l^\prime_i\end{array}\right)_{L}: \left(1/2:0:-1\right),~
L_{iR}=\left( \begin{array}{c}
\nu^\prime_i\\l^\prime_i\end{array}\right)_{R}: \left(0:1/2:-1\right) \ , 
\nonumber \\ Q_{iL}&=&\left( \begin{array}{c}
u^\prime_i\\d^\prime_i\end{array}\right)_{R}: \left(1/2:0:1/3\right),~
Q_{iR}=\left( \begin{array}{c}
u^\prime_i\\d^\prime_i\end{array}\right)_{R}: \left(0:1/2:1/3\right) \ . 
\eeqa
Here, $i=1,2,3$ stands for generation number, and $(I_L, I_R, Y\equiv
B-L)$ labels representation of the gauge group for each multiplet. The
representation determines interactions of the multiplet with gauge
fields. Before spontaneous symmetry breaking (SSB) the latter include
$W^{a,\mu}_{L}$, $W^{a,\mu}_{R}$ ($a=1,2,3$), and $B^{\mu}$ for $\rm
SU(2)_L$, $\rm SU(2)_R$, and $\rm U(1)_{B-L}$ gauge group factors,
respectively.

The SSB is achieved via the Higgs mechanism. The Higgs sector of the
theory is not unique. However, the main results of this paper are
largely independent of the details of the Higgs sector provided the
LRSM has triplet Higgses and therefore heavy right-handed neutrinos.
In our study  we  
choose~\cite{Mohapatra:1979ia,Mohapatra:1980yp} a Higgs
sector that consists of the bi-doublet $\phi:(1/2,1/2,0)$ and two triplets
$\Delta_L:(1,0,2)$ and $\Delta_R:(0,1,2)$:
\beqa \phi&=&\left( \begin{array}{cc} \phi_1^0 & \phi_2^+\\\phi_1^- &
\phi_2^0\end{array}\right),~ \Delta_{L,R}=\left( \begin{array}{cc}
\delta_{L,R}^+/\sqrt{2} & \delta_{L,R}^{++} \\ \delta_{L,R}^0 &
-\delta_{L,R}{^+}/\sqrt{2}
\end{array}\right) \ , \\
\langle\phi\rangle&=&\left( \begin{array}{cc} \kappa_1/\sqrt{2} & 0\\
0 & \kappa_2/\sqrt{2}\end{array}\right),~
\langle\Delta_{L,R}\rangle=\left( \begin{array}{cc} 0 & 0
\\
v_{L,R} & 0 \end{array}\right) \ , 
\eeqa
where the vacuum expectation values (VEVs) are shown in the second
line. The most general Higgs potential with this field content has
been analyzed in Ref.~\cite{Deshpande:1990ip}.  If one requires the
scale $v_R$ in the multi-TeV range (but not significantly larger), the
only choice which avoids excessive fine-tuning and leads to acceptable
phenomenology is to set to zero certain 
couplings in the Higgs potential as well as $v_L$~\cite{Deshpande:1990ip}. 
Moreover, we assume no explicit or spontaneous CP violation 
in the Higgs sector~\cite{Barenboim:2001vu}.   
In summary, two distinct mass scales appear in the model: the
electroweak symmetry breaking scale $\kappa \sim \kappa_1 \sim
\kappa_2 \sim 250$ GeV, and the scale $v_R$ at which $SU(2)_R$ and $U
(1)_{B-L}$ are spontaneously broken.  Phenomenological considerations  
require $v_R \gg \kappa$.

\subsection{Physical fields}

After SSB matter and gauge fields acquire non-vanishing masses, 
which generally allow for mixing of the fields with the same quantum
numbers. In the following, we identify masses and mixing angles which
are important for our calculation. We omit the discussion
of the quark sector of the model, as it is irrelevant for our work. 

\subsubsection{Leptons}

The $3\times 3$ mass matrix for charged leptons is
$M_l=\left(y_D\kappa_2+{\tilde y_D \kappa_1}\right)/\sqrt{2}$, where
$y_D$ and ${\tilde y_D}$ are, respectively, the Yukawa coupling
matrices for the bi-doublet $\phi$ and its charge conjugate. 
$M_l$ is diagonalized by a biunitary transformation
$V_L^{l\dagger}M_lV_R^l=(M_l)_{diag}$. 
Since $(M_l)_{diag} \ll \kappa$, one has $y_D, \tilde{y}_D \ll 1$. 
Here, $V_{L,R}^{l}$ are $3\times 3$
unitary matrices. These matrices relate the charged lepton mass
eigenstates $l_{L,R}$ to the corresponding flavor eigenstates from
Eq.~(\ref{eq:matter}): $l_{L,R}^\prime=V_{L,R}^l l_{L,R}$.

Within the LRSM it is convenient to describe neutrino fields by
four-component spinors
\beq
n_R^\prime=\left( \begin{array}{c}
\nu_R^{\prime c}
\\
\nu_R^\prime
\end{array}\right),~
n_L^\prime=\left( \begin{array}{c}
\nu_L^{\prime}
\\
\nu_L^{\prime c}
\end{array}\right),~
\nu_{L,R}^{\prime c}=i\sigma_2\nu_{R,L}^{\prime *} \ . 
\eeq
The $6 \times 6$ mass matrix for the neutrinos is of see-saw type.
It has both Majorana and Dirac entries:
\beq
M_\nu=\left( \begin{array}{cc}
0 & M_D
\\
M_D^T & M_R
\end{array}\right),~M_D
={1\over \sqrt{2}}\left(y_D\kappa_1+{\tilde y_D \kappa_2}\right),~
M_R=\sqrt{2}y_M v_R \ , 
\eeq
where $y_M$ is a $3\times 3$ Majorana-type Yukawa coupling
matrix. $M_\nu$ is diagonalized by a $6\times 6$ unitary matrix $V$:
$V^TM_\nu V=(M_\nu)_{diag}$. This matrix relates neutrino mass and
flavor eigenstates: $n_L^\prime=V^* N_L$, $n_R^\prime=V N_R$. 
Three eigenvalues of $M_\nu$ are small (denoted later by $m_\nu$), of the
order of $M_D^2/M_R$, and correspond to the light neutrinos of the SM.  
For $M_R$ in the multi-TeV range, identifying $M_D$ with the charged
lepton mass clearly violates the 95\% C.L. limit $\sum m_\nu < 0.7
$~eV~\cite{Spergel:2003cb} from WMAP.  However, appropriate choices of
$y_D$ and $\tilde{y}_D$ (leading to $M_D$ with non-trivial family
structure and absolute scale on the order of $m_e/\kappa \sim 10^{-6}$) 
result into $m_{\nu_i}$ consistent with present phenomenology. 
The remaining three eigenstates are predominantly right-handed with
mass $M_n \sim M_R\sim y_M v_R\gsim$ 1 TeV, since we assume $y_M \sim
{\cal O} (1)$.  The amount of heavy-light mixing of the neutrino
sector is set by the ratio $\epsilon\sim M_D/M_R\sim y_D\kappa/y_M v_R
\ll 1$. As discussed below, we will expand our results in $\epsilon$
and $\kappa / v_R$, and retain leading non-vanishing order.

The LFV couplings of leptons to gauge
and Higgs bosons are conveniently parameterized in terms of two
$6\times 3$ matrices \cite{Duka:1999uc}
\beqa
K_L=V_L^{\nu \dagger}V_L^l,~
K_R=V_R^{\nu \dagger}V_R^l,~
V=\left( \begin{array}{c}
V_L^{\nu *}
\\
V_R^\nu  
\end{array}\right) \ . 
\eeqa
At the leading order in $\epsilon$, the upper $3\times3$ block of 
$K_L$ and the lower $3\times3$ block of $K_R$, respectively,
describe flavor mixing in the light and heavy neutrino 
sectors. They are analogous to the CKM matrix which appears in the 
quark sector of the SM, and satisfy unitarity conditions 
up to corrections of order $\epsilon^2$. 
In particular, 
the upper $3\times 3$ block of $K_L$ is the familiar mixing matrix for 
light neutrinos \cite{Hagiwara:fs}. 
As observed in the introduction, 
contributions involving light neutrinos (and
$K_L$) to any LFV process are GIM suppressed relative to those involving
heavy neutrinos. 
Therefore, the leading contributions to the LFV processes we consider
depend on the masses and flavor mixing of heavy neutrinos only.

\subsubsection{Gauge fields}

The charged gauge bosons acquire the following mass matrix
\beq
{\tilde M}_W^2={g^2\over 4}\left( \begin{array}{cc}
\kappa_+^2 & -2\kappa_1\kappa_2
\\
-2\kappa_1\kappa_2 & \kappa_+^2+2v_R^2
\end{array}\right) ~,
\eeq
which is diagonalized via the mixing angle
$\xi=-\tan^{-1}\left(2\kappa_1\kappa_2/v_R^2\right)/2$ with the
eigenvalues $M_{W_{1,2}}^2=g^2\left(\kappa_+^2+v_R^2 \mp
\sqrt{v_R^4+4\kappa_1^2\kappa_2^2}\right)/4$. Here
$\kappa_+=\sqrt{\kappa_1^2+\kappa_2^2}$, and the mass eigenstates 
are related to the gauge eigenstates by 
\beq
W_L = \cos \xi \, W_1  + \sin \xi \, W_2, \qquad
W_R = - \sin \xi \, W_1  + \cos \xi \, W_2 \ . 
\eeq 

The mass matrix for the neutral gauge bosons is
\beq {\tilde M}_0^2={1\over 2}\left( \begin{array}{ccc} {g^2\over
2}\kappa_+^2 & -{g^2\over 2}\kappa_+^2 & 0 \\ -{g^2\over 2}\kappa_+^2
& {g^2\over 2}\left(\kappa_+^2+4v_R^2\right) & -2gg^\prime v_R^2 \\ 0
& -2gg^\prime v_R^2 & 2g^{\prime 2} v_R^2
\end{array}\right)
\eeq
It has the following non-zero eigenvalues  (the third one is vanishing)
\beqa
M_{Z_{1,2}}^2&=&{1\over 4}\Biggl[g^2\kappa_+^2+2v_R^2(g^2+g^{\prime
2})\nonumber \\ 
&\mp& \sqrt{\left[g^2\kappa_+^2+2v_R^2(g^2+g^{\prime
2})\right]^2-4g^2(g^2+2g^{\prime 2})\kappa_+^2v_R^2}\Biggr] 
\eeqa 
The explicit form of the unitary matrix that diagonalizes ${\tilde
M}_0^2$ is given in Ref.~\cite{Duka:1999uc}.  Here, we only list the
expression for the $Z_1-Z_2$ mixing angle:
\beq \phi=-{1\over 2}\sin^{-1}{g^2\kappa_+^2\sqrt{\cos2\theta_W}\over
2 c_W^2(M_{Z_2}^2-M_{Z_1}^2)} \eeq
where $\theta_W$ is the weak mixing angle, 
and $c_W(s_W)$ is $\cos \theta_W (\sin \theta_W)$
(we use this abbreviation throughout). 
In this paper we work in the
regime where $\phi,\xi\ll 1$ \footnote{The experimental limits
on the gauge boson mixing angles are 
$| \xi | < 3 \times 10^{-3}$ and $|\phi |
< 1.8 \times 10^{-3}$~\cite{Hagiwara:fs}}, 
since both are ${\cal O}( (\kappa/v_R)^2)$. 
Note that because $y_D\ll y_M$ we have $\epsilon < \phi,\xi$.

\subsubsection{Higgs fields}

With the Higgs sector described above, there are six neutral and four charged
physical Higgs bosons \cite{Duka:1999uc}.   
At leading order in $\epsilon$, however, the neutral Higgs bosons do not
contribute to the LFV processes involving charged leptons in the
external states, and we do not consider them in the following. Two of
the remaining bosons, $H_{1,2}^+$, are singly charged, with masses
$M_{H_{1,2}}$. The last two bosons, $\delta_{L,R}^{++}$, are doubly charged, 
with masses $M_{\delta_{L,R}^{++}}$. 
Masses of the Higgs bosons depend on a number of parameters in the
Higgs potential, with the natural scale 
$M_H \sim M_{\delta} \sim v_R$ \cite{Duka:1999uc,Deshpande:1990ip}.

\subsection{Lepton Interactions}

The LFV interactions of leptons with gauge ($W_2$),    
singly and doubly charged  bosons are given by the following 
lagrangian densities:
\bea
{\cal L}_{CC} &=& \displaystyle\frac{g}{\sqrt{2}} \Bigg\{
\overline{N} \Big[ \gamma^\mu \, P_R \, (K_R) \Big] l  \cdot
W_{2 \ \mu}^+
+
\overline{l} \Big[ \gamma^\mu \, P_R \, (K_R^\dagger) \Big] N \cdot
W_{2 \ \mu}^-
\Bigg\}  \\
{\cal L}_{H_1
} &=& \displaystyle\frac{g}{\sqrt{2}} \ \Bigg[
H_{1}^+  \
\overline{N}  \left( \tilde{h} \, P_L
\right) l
+ H_1^-  \
\overline{l}  \left( \tilde{h}^\dagger  \, P_R
\right) N
\Bigg]
\label{eq:xlag1}  \\
{\cal L}_{\delta^{\pm \pm}_{L,R}} &=&
\displaystyle\frac{g}{2} \ \Bigg[
\delta_{L,R}^{++}  \
\overline{l^{c}}  \left( h_{L,R} \, P_{L,R} \right) l      +
\delta_{L,R}^{--}  \
\overline{l}  \left( h_{L,R}^{\dagger} \, P_{R,L} \right) l^{c}
\Bigg]
\label{eq:xlag2} \  , 
\eea
where  $P_{L,R} = (1 \mp \gamma_5)/2$,  $N=N_L+N_R=N^c$, $l=l_L+l_R$
and where we have neglected ${\cal O}(\xi)$ terms.
With our choice of the Higgs sector, it follows~\cite{Deshpande:1990ip} 
that  the $3 \times 3$ matrix couplings $h_L$ and $h_R$ can 
be identical (manifest left-right symmetry) or can have components 
differing  by a sign (quasi-manifest left-right symmetry). 
In the manifest left-right symmetry case one finds 
\beq
h_L = h_R = K_R^T \, \frac{M_{\nu}^{\rm diag}}{M_{W_2}} \, K_R \equiv h 
\qquad \qquad \tilde{h} = K_L^* \, h_L \ . 
\eeq
Note that it is $K_R^T$, not $K_R^\dagger$ that appears in the
definition of $h$. Because $K_R$ may contain Majorana phases, 
$h$ is not necessarily proportional to the unit matrix even if all
heavy neutrinos are degenerate.  At leading order in $\epsilon $ one
has
\beq 
h_{i j} = \sum_{n={\rm heavy}} \, \Big( K_R \Big)_{n i}  
\Big( K_R \Big)_{n j}  \, \sqrt{x_n}  \ , 
\label{eq:hcouplings1}
\eeq
\beq
\Big( h^\dagger  h \Big)_{e \mu} =
\Big( \tilde{h}^\dagger  \tilde{h} \Big)_{e \mu} =
\displaystyle\sum_{n={\rm heavy}}
x_n  \Big( K_R^\dagger \Big)_{e n}
\Big( K_R \Big)_{n \mu}   \equiv g_{\rm lfv} \ ,
\label{eq:xcouplings}
\eeq
\beq
 x_n = \left({M_n\over M_{W_2}}\right)^2\ \ \ , 
\eeq
where the sum is over the heavy neutrinos only.
Eq.~(\ref{eq:xcouplings}) relates the lepton-gauge boson couplings to
the lepton-Higgs triplet couplings. We emphasize that it is specific
to left-right symmetric models, and plays a central role in
phenomenological applications. 
Generalization to the quasi-manifest left-right symmetry case  
is trivial and we have explicitly checked that a possible  
relative sign between $h_L$ and $h_R$ has no observable 
consequences in LFV processes.
Finally, note that for degenerate
heavy neutrinos, i.e., $x_n = const$, 
one has $g_{\rm lfv} = {\cal O}(\epsilon^2)$ due to the 
approximate unitarity of the lower $3\times3$ block of  
$K_R$. Thus $g_{\rm lfv}$ depends only on the mass square differences
of the heavy neutrinos. The same is not true for the individual $h_{ij}$s.

\section{Calculation}

\begin{figure}[ht]
\caption{ 
Diagrams contributing to $\mu \rightarrow e \gamma$, $\mu
\rightarrow e $ conversion in nuclei, and $\mu \rightarrow 3 e$. 
The wavy lines represent neutral gauge bosons ($\gamma$ or $Z_{1,2}$). 
$\mu \rightarrow e \gamma$ is described by class (a). 
$\mu \rightarrow e $ conversion is described by class (a) 
(attaching a quark line to the neutral gauge boson) and  
class (b) (with two external quark legs). 
$\mu \rightarrow 3 e$ receives in principle contributions 
from classes (a) (attaching an electron line to the gauge boson), 
(b), and (c). }
\label{fig:fig1}
\centering
\begin{picture}(100,200)  
\put(0,10){\makebox(100,20){\epsfig{figure=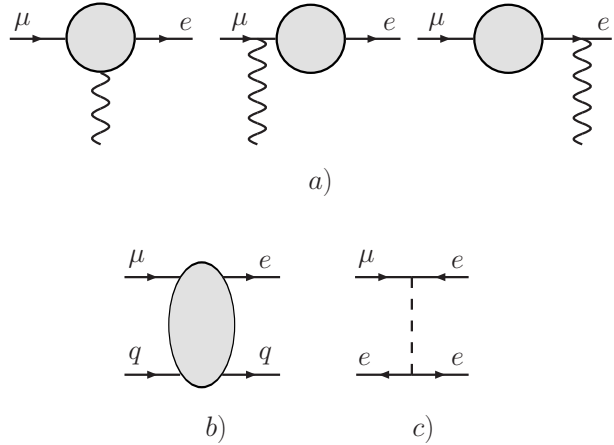,width=9cm}}}
\end{picture}
\end{figure}

Within the LRSM we performed a complete calculation of the LFV  
muon processes  $\mu \rightarrow e \gamma$, $\mu
\rightarrow e $ conversion in nuclei, and $\mu \rightarrow 3 e$ to 
leading order in the expansion parameters 
$\kappa/v_R$ and $\epsilon \sim M_D /M_R$. 
Diagrammatic contributions fall into three classes, 
schematically shown in Fig.~\ref{fig:fig1}.   
Given the lagrangian in the physical fields basis~\cite{Duka:1999uc},
we have first identified all diagrams contributing to leading order in
$\kappa/v_R$ and $M_D /M_R$. 
We have then calculated the LFV vertices $\mu \rightarrow e \gamma^*$,
$\mu \rightarrow e Z_1^*$, and $\mu \rightarrow e Z_2^*$
[Fig.~\ref{fig:fig1} (a)].  Finally, we have combined the LFV
effective vertices with the $\bar{q} q \gamma^*$, $\bar{q} q
Z_{1,2}^*$ interactions, and relevant box-type [Fig.~\ref{fig:fig1}
(b)] and tree-level diagrams [Fig.~\ref{fig:fig1} (c)] to obtain the
effective lagrangian for $\mu \rightarrow e $ conversion and $\mu
\rightarrow 3 e$.  Our calculation and main results are described in
this section, and some technical details are given in the
appendixes~\ref{app:functions}, \ref{app:mueg}.

\subsection{Identifying the leading contributions}
In the LRSM at low energy all effects of the right-handed sector are
suppressed by powers of $\kappa / v_R$, as a consequence of the
decoupling theorem~\cite{Appelquist:tg}.  In our analysis we keep only
the leading contributions in the expansion parameters $\kappa/v_R$ and
$\epsilon$.  Throughout, we work in 't Hooft-Feynman gauge.  Our findings
can be summarized as follows:

\begin{itemize}

\item The $\mu \rightarrow e \gamma^*$ vertex receives its leading
contributions at order $\epsilon^0,(\kappa/v_R)^0$. 
In accordance with electromagnetic gauge invariance, however, the
momentum-independent piece of the amplitude vanishes, so that the
resulting vertex function is actually of order $q^2/v_R^2$, $q$ being
the momentum transfer.  Our expression is fully gauge invariant and
respects the decoupling theorem. 
Note that when this vertex is inserted into the $\mu\to e$ conversion
amplitude, the $q^2$ in the vertex cancels the $1/q^2$ of the
photon propagator, leaving a contact interaction that scales  
as $\sim 1/v_R^2$.

\item  The $\mu  \rightarrow e Z_2^*$ vertex is again not
suppressed by powers of $\epsilon$ or $(\kappa/v_R)$,
and we only keep the momentum-independent component. This result does not 
contradict the decoupling theorem, as one of the external states belongs 
to the heavy sector of the theory.  
Since $q^2 << M_{Z_2}^2$, the contribution from this  
vertex to the $\mu\to e$ conversion amplitude goes as $1/v_R^2$.

\item The $\mu \rightarrow e Z_1^*$ vertex nominally receives its
leading contributions at order $\epsilon^0,(\kappa/v_R)^0$.  However, the
momentum-independent part of this class of diagrams sums to zero, in
accordance with the decoupling theorem.  We find that the leading
non-vanishing contribution is ${\cal O} ( (\kappa/v_R)^2 )$.
Consequently, the contribution from this vertex to the  
conversion amplitude is also $\sim 1/v_R^2$. 

\end{itemize}

For the kinematics of the LFV decays considered here, the momentum 
dependent contributions to the  $\mu \rightarrow e Z_{1,2}^*$ vertices 
are highly suppressed and can be neglected. 

\subsection{Effective vertices}
The $\mu \rightarrow e$ LFV vertices can be expressed in terms of 
known couplings and form factors $F_{L,R}^{(i)}, A_{L,R}$  
as follows, 
\bea
L_{\mu}^{(Z_1)} &=& \displaystyle\frac{e \, G_F M_{W_1}^2}{\sqrt{2} (4 \pi)^2} 
\, \displaystyle\frac{1}{s_W \, c_W} \, 
\overline{e} \gamma_\mu   
\left(F_L^{(1)} P_L  + F_R^{(1)} P_R  \right) \mu    \ , 
\\ 
L_{\mu}^{(Z_2)} &=& \displaystyle\frac{e \, G_F M_{W_1}^2}{\sqrt{2} (4 \pi)^2} 
\, \displaystyle\frac{1}{s_W \, c_W 
\sqrt{\cos 2 \theta_W}} \, 
\overline{e} \gamma_\mu  
\left(F_L^{(2)} P_L  + F_R^{(2)} P_R  \right) \mu    \ , 
\\
L_{\mu}^{(\gamma)} &=& \displaystyle\frac{e \, G_F}{\sqrt{2} (4 \pi)^2} \, 
\overline{e} \left\{ \left(q^2 \gamma_\mu -  / \! \! \!q q_\mu \right)
\left(F_L^{(\gamma)} P_L  + F_R^{(\gamma)} P_R  \right) 
\right. \nonumber \\
&  &  \left. - i 8 (4 \pi)^2  m_\mu  \sigma_{\mu \nu} q^\nu 
\left(A_L  P_L  + A_R P_R  \right) 
\right\} \mu \ \ ,  
\label{eq:photon}
\eea
where $q = p_e - p_\mu$, $\sigma_{\mu \nu} = \frac{i}{2} [ \gamma_\mu
, \gamma_\nu ]$.  The $\mu \, e \, \gamma^*$ effective
vertex has both ``anapole'' ($F_{L,R}^{(\gamma)}$) 
and dipole ($A_{L,R}$) terms 
\footnote{The first term in Eq.~(\ref{eq:photon}) involves a
coupling of the flavor-violating lepton current to the 
electromagnetic current rather than to a field associated with the 
corresponding vector potential. Zeldovich referred to this interaction as an 
anapole coupling~\cite{Zeldovich}.}.
Only the dipole terms contribute to the on-shell decay $\mu
\rightarrow e \gamma$, while both anapole and dipole contribute to
$\mu$ to $e$ conversion in nuclei.

\begin{figure}[!bh]
\caption{Basic topologies for the 
one-particle-irreducible contributions to 
the $\mu \, e \, \gamma^*$, $\mu \, e \, Z_1^*$, and 
$\mu \, e \, Z_2^*$ effective vertices (detailed version of 
Fig.~\ref{fig:fig1} (a)).  
Wavy lines represent gauge bosons, dashed lines represent scalars 
(physical or unphysical), full lines represent leptons (charged or 
neutral). The internal particles contributing at leading order 
to  each topology are listed in Table \ref{tab:table2}. 
}
\label{fig:vertices}
\centering
\begin{picture}(200,270)  
\put(45,100){\makebox(100,20){\epsfig{figure=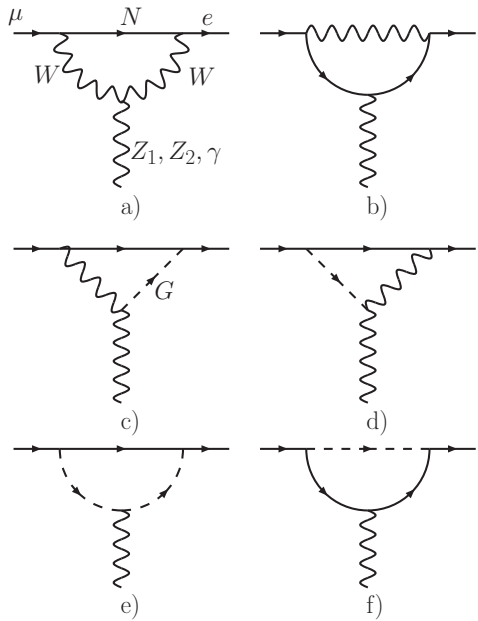,height=11.5cm}}}
\end{picture}
\end{figure}

The $\mu \rightarrow e$ effective vertices receive contributions from
the one-particle-irreducible diagrams depicted in
Fig. \ref{fig:vertices}, as well as from external-leg corrections.
The vertex corrections can be grouped into three classes: (i) gauge
contributions (including unphysical Higgs exchange), (ii) singly
charged physical Higgs contributions, and (iii) doubly charged Higgs
contributions.
Power counting implies that only certain combinations of gauge bosons,
neutrinos, and Higgs particles contribute to leading order in
$\kappa/v_R$ and $\epsilon$. The relevant intermediate states are indicated
diagram by diagram in Table \ref{tab:table2}.

\begin{table*}[hb]
\caption{\label{tab:table2}
Intermediate states contributing at leading order in $\kappa/v_R$
and $y_D$ to $\mu \, e \, \gamma^*$, $\mu \, e \, Z_1^*$, and 
$\mu \, e \, Z_2^*$
effective vertices in 't Hooft-Feynman gauge. 
For each topology in Fig.\ref{fig:vertices} we
list the intermediate states as they appear starting from the 
muon vertex and following the loop counter-clockwise.
Neutrinos are denoted by $N_{\rm h}$ (heavy) and $N_{\rm l}$ (light). 
$G_{1,2}$ denote the unphysical Higgs fields associated with the longitudinal 
polarization of the gauge bosons $W_{1,2}$.
}
\begin{ruledtabular}
\begin{tabular}{c|c|c|c|c|c|c}
 & a) & b) & c) & d) & e) & f) 
\\
\hline
$\gamma$  &  $W_{2},W_{2}, N_{\rm h}$  &    & 
$W_2,G_2,N_{\rm h}$   & $G_2,W_2,N_{\rm h}$    
& $G_2,G_2,N_{\rm h}$   &  
\\
 &   &   &   &   & $H_1,H_1,N_{\rm l}$   &  
\\
 &   &   &   &   &  $\delta_{L,R}^{\pm \pm},\delta_{L,R}^{\pm \pm},l_i$ 
  &  $l_i, l_i, \delta_{L,R}^{\pm \pm} $
\\
\hline
$Z_1$  &  $W_{2},W_{2}, N_{\rm h}$  & $N_{\rm h}, N_{\rm h}, W_2$ 
& 
$W_2,G_2,N_{\rm h}$   & $G_2,W_2,N_{\rm h}$    & 
$G_2,G_2,N_{\rm h}$   &  $N_{\rm h},N_{\rm h},G_2$    
\\
      &     &     & 
$W_2, H_2, N_{\rm h}$   & $H_2,W_2,N_{\rm h}$ 
 & $H_2,H_2,N_{\rm h}$ ;   $G_2,H_2,N_{\rm h}$ ; 
$H_2,G_2,N_{\rm h}$    &      
\\
      &     &     & 
$W_2, G_1, N_{\rm h}$   & $G_1,W_2,N_{\rm h}$ 
 & $G_1,G_1,N_{\rm h}$ ;   $G_1,G_2,N_{\rm h}$ ; 
$G_2,G_1,N_{\rm h}$    &      
\\
\hline
$Z_2$  &  $W_{2},W_{2}, N_{\rm h}$  & $N_{\rm h}, N_{\rm h}, W_2$
  & 
$W_2,G_2,N_{\rm h}$   & $G_2,W_2,N_{\rm h}$  
  & $G_2,G_2,N_{\rm h}$   &  $N_i,N_i,G_2$    
\\ 
\end{tabular}
\end{ruledtabular}
\end{table*}

\subsubsection{$\mu \, e \, Z_1^*$ vertex}  
In this case the leading diagrams involving triplet Higgs (singly
and doubly charged) sum to zero, and the main effect stems from gauge
contributions.  When working in 't Hooft-Feynman gauge, one needs to
include the effect of unphysical Higgs exchange, and their mixing with
other physical and unphysical scalars of the theory 
(terms proportional to $\eta_1$ and $\eta_2$ below).
In terms of the heavy neutrino masses ($M_n$), mixing matrix $K_R$,
and the ratios $x_n = (M_n/M_{W_2})^2$, $y_n = (M_n/M_{W_1})^2$, 
$z_n = (M_n/M_{H_2})^2$, the 
resulting form factors have the following structure:
\bea
F_R^{\rm (1)} & = & 
\displaystyle\sum_{n={\rm heavy}} \, \Big( K_R^\dagger \Big)_{e n} 
\Big( K_R \Big)_{n \mu}  \ \, 
\Bigg[  
\eta_0 
\   S_1 (x_n) +   
2  \,  \eta_1  D_{1} (x_n, y_n)  + \eta_2 D_{1} (x_n, z_n) 
\Bigg]    \ , \\
F_L^{\rm (1)} & = & {\cal O} \left( 
\displaystyle\frac{m_\nu^2}{M_{W_1}^2} \right) \ll F_R^{\rm (1)} 
\ ,  
\eea
where  ($\kappa_-^2 = \kappa_1^2 - \kappa_2^2 $)  
\bea
\eta_0 &=&  - \displaystyle\frac{\sin \phi \ c^2_W}{
\sqrt{\cos 2 \theta_W}} \simeq 
\displaystyle\frac{M_{Z_1}^2 \, c^2_W}{M_{Z_2}^2 - M_{Z_1}^2} 
\simeq 
\displaystyle\frac{1 - 2  s^2_W}{2 c^2_W} 
\frac{M_{W_1}^2}{M_{W_2}^2}  \ , 
\\
\eta_1 &=& \left(\frac{\kappa_1 \kappa_2 } {\kappa_+ v_R}\right)^2 
\simeq \frac{1}{2} \left( \frac{M_{W_2}}{M_{W_1}} \, 
\sin \xi \right)^2 \leq {1\over 2}\left({M_{W_1}\over M_{W_2}}\right)^2 \ , 
\\
\eta_2 &=& \left(\frac{\kappa_-^2}{\sqrt{2}\kappa_+ v_R}\right)^2 \leq 
\left({M_{W_1}\over M_{W_2}}\right)^2 \ ,  
\eea
and the functions $S_{1}(x),D_1 (x,y)$ 
are defined in Appendix~\ref{app:functions}.

\subsubsection{$\mu \, e \, Z_2^*$ vertex}  
As in the previous case, the leading term arises from gauge-lepton
interactions, and the leading physical Higgs effects cancel out.
With the notation established above, we find:
\bea
F_R^{\rm (2)} & = & 
c^2_W 
\displaystyle\sum_{n={\rm heavy}} \, \Big( K_R^\dagger \Big)_{e n} 
\Big( K_R \Big)_{n \mu}  \ \, 
S_1 (x_n)  \ ,  \\
F_L^{\rm (2)} & = & {\cal O} \left( 
\displaystyle\frac{m_\nu^2}{M_{W_1}^2} \right) \ll F_R^{\rm (2)} 
\ . 
\eea

\subsubsection{$\mu \, e \, \gamma^*$ vertex}

Both anapole and dipole transition form factors receive non-vanishing
leading contribution from gauge diagrams and exchange of singly and
doubly charged triplet Higgs particles.  Neglecting charged fermion
masses (see Appendix~\ref{app:mueg}) and using Eq.~(\ref{eq:xcouplings}),  
the various amplitudes read: 
\bea
F_R^{\rm (\gamma)} & = & 
\displaystyle\sum_{n={\rm heavy}} \, \Big( K_R^\dagger \Big)_{e n} 
\Big( K_R \Big)_{n \mu}  \ 
\left[ \frac{M_{W_1}^2}{M_{W_2}^2}  \, S_2 (x_n) 
- x_n \, \frac{8}{3}  
\frac{M_{W_1}^2}{M_{\delta_R^{++}}^2} \, \log 
\left( \frac{-q^2}{M_{\delta_R^{++}}^2} \right) 
\right] \ , 
\\
F_L^{\rm (\gamma)} & = & 
\displaystyle\sum_{n={\rm heavy}} \, \Big( K_R^\dagger \Big)_{e n} 
\Big( K_R \Big)_{n \mu}  \  \, x_n 
\left[  - \frac{8}{3}  \frac{M_{W_1}^2}{M_{\delta_L^{++}}^2}   \, 
\log \left( \frac{-q^2}{M_{\delta_L^{++}}^2} \right)  
- \frac{2}{9} \frac{M_{W_1}^2}{M_{H_1^+}^2}  
\right] \ , 
\\
A_L & = & \frac{1}{16 \pi^2} \, 
\displaystyle\sum_{n={\rm heavy}} \, \Big( K_R^\dagger \Big)_{e n} 
\Big( K_R \Big)_{n \mu}  \ 
\left[ \frac{M_{W_1}^2}{M_{W_2}^2}  \, S_3 (x_n) 
-  \frac{x_n}{3}  \frac{M_{W_1}^2}{M_{\delta_R^{++}}^2}  
\right] \ , 
\\
A_R & = & \frac{1}{16 \pi^2} \, 
\displaystyle\sum_{n={\rm heavy}} \, \Big( K_R^\dagger \Big)_{e n} 
\Big( K_R \Big)_{n \mu}  \ \, x_n 
\left[  -  \frac{1}{3}  \frac{M_{W_1}^2}{M_{\delta_L^{++}}^2}  
-  \frac{1}{24}  \frac{M_{W_1}^2}{M_{H_1^+}^2}  
\right] \ . 
\eea
The functions $S_{2,3}(x)$ are given explicitly in 
Appendix~\ref{app:functions}.  
The most important feature of these results is the logarithmic
enhancement ($q^2 \simeq -m_\mu^2$) of the anapole transition form
factors, arising from the doubly charged triplet Higgs diagrams.  This
implies that in the left-right symmetry framework, $\mu \rightarrow e$
conversion in nuclei is as strong  probe of LFV as $\mu \rightarrow e \gamma$
since its amplitude is logarithmically enhanced,
and thus compensates for the extra factor of $\sim \alpha$. 
This effect was pointed out for a larger class of models in 
Ref.~\cite{Raidal:1997hq} within an effective field theory
approach. Its consequences within the LRSM  will be
discussed in the next section in detail.

In Appendix \ref{app:mueg} we report full expressions for the $\mu \,
e \, \gamma^*$ form factors (including charged fermion masses) in
terms of $h,\tilde{h}$ (i.e. without using Eq.~(\ref{eq:xcouplings})).

\subsection{Effective Lagrangian for $\mu \rightarrow e$ conversion}

\begin{figure}[!t]
\centering
\begin{picture}(200,100)  
\put(45,-75){\makebox(100,20){\epsfig{figure=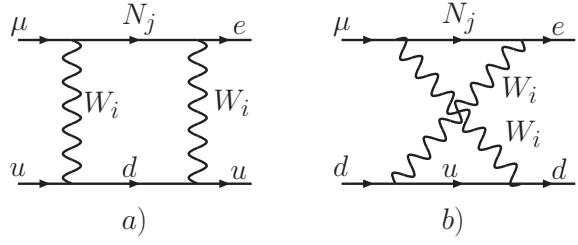,width=9.5cm}}}
\end{picture}
\caption{Box diagrams contributing to $F_R^{(\rm B)}$ 
(Fig.~\ref{fig:fig1} (b)).}
\label{fig:boxes}
\end{figure}

The effective Lagrangian for $\mu \rightarrow e $ conversion receives
contributions from (i) tree level exchange of heavy neutral Higgs
states; (ii) box diagrams depicted in Fig.\ref{fig:boxes}; (iii) LFV
effective vertices, with the gauge boson attached to a quark line (the
relevant quark-gauge couplings are summarized in Table
\ref{tab:table1}).  Inspection of the neutral Higgs couplings implies
that the ratio of effective couplings $g_{\rm tree}^{\rm eff}$ and
$g_{\rm loop}^{\rm eff}$ generated by tree level Higgs exchange and
loop corrections, respectively, scales as $g_{\rm tree}^{\rm
eff}/g_{\rm loop}^{\rm eff} \sim (y_D)^2 / (\alpha/4 \pi) \ll 1 $.
Therefore, we safely neglect the Yukawa suppressed tree level
diagrams.

After casting the $\mu \, e \, Z_{1,2}^*$ vertices and the $\mu \, e \,
\gamma^*$ anapole terms in the form of a current-current interaction,
the effective lagrangian can be written as in
Ref.~\cite{Kitano:2002mt}:
\bea
{\cal L}_{\mu \to e} &=& 
- \displaystyle\frac{4  G_F \, e}{\sqrt{2}}   
\ m_\mu \   \overline{e} \, 
\sigma_{\mu \nu} (A_L P_L + A_R P_R) \mu 
\cdot F^{\mu \nu}   
\nonumber \\ 
& &  - \displaystyle\frac{G_F}{\sqrt{2}}  \displaystyle\sum_{q} \, 
\Bigg\{  
\overline{e} \gamma_\mu  \Big[ g_{LV} (q) P_L + 
g_{RV}(q) P_R \Big] \mu  \otimes \overline{q} \gamma^\mu q 
\nonumber \\
& &  \qquad 
+ \overline{e} \gamma_\mu  \Big[ g_{LA} (q) P_L + 
g_{RA}(q) P_R \Big] \mu  \otimes \overline{q} \gamma^\mu \gamma^5 q 
\Bigg\} + {\rm h.c.} \ ,  
\eea 
where $F^{\mu \nu}$ has to be understood as the classical field 
produced by the nucleus.  
In terms of the box contribution ($S_4(x)$ is defined in 
Appendix~\ref{app:functions}) 
\bea
F_R^{\rm (B)} & = &  8 \, 
\displaystyle\sum_{n={\rm heavy}} \, \Big( K_R^\dagger \Big)_{e n} 
\Big( K_R \Big)_{n \mu}  \ \, S_4 (x_n)  \  , 
\eea 
and the LFV form factors, the couplings $g_{LV,RV} (q)$ are  
\bea
g_{LV}(q) &=& - \displaystyle\frac{\alpha}{4 \pi} \,
 F_{L}^{(\gamma)} \, v_{q}^{(\gamma)}  \ , 
\\ 
g_{RV}(q) &=&   \frac{\alpha}{8 \pi s_W^2} \Bigg\{ 
- 2 \sin^2 \theta_W   F_{R}^{(\gamma)} \, v_{q}^{(\gamma)} 
+ \displaystyle\frac{1}{2} F_R^{(1)} v_q^{(1)}  
\nonumber \\
 & & 
+ \displaystyle\frac{M_{W_1}^2}{M_{W_2}^2}  
\displaystyle\frac{F_R^{(2)} \, v_q^{(2)} 
}{4 c_W^4 } \, 
- 
\displaystyle\frac{M_{W_1}^2}{M_{W_2}^2}  
F_R^{(B)}  v_q^{(B)}  
\Bigg\}  \ . 
\label{eq:grv}
\eea
The expressions for $g_{LA,RA} (q)$ are obtained by replacing
$v_q^{(i)}$ with $a_q^{(i)}$ in $g_{LV,RV} (q)$. We remark that all
the contributions to $g_{RV}(q)$ in Eq.~(\ref{eq:grv}) enter at
leading order $\kappa /v_R$, contrary to what appears in earlier
calculations~\cite{Riazuddin:hz,Barenboim:1996vu}.  In
Ref.~\cite{Riazuddin:hz} only $F_R^{(2)}$ and $F_R^{(B)}$ were
included, while the authors of Ref.~\cite{Barenboim:1996vu} considered
only $F_{L,R}^{(1)}$. Both of these previous studies omitted the
dominant, logarithmically-enhanced contributions from
$F_{L,R}^{(\gamma)}$.
Finally,  we note that upon taking matrix elements
of ${\cal L}_{\mu \to e}$ in nuclei, the following combinations of
$g_{LV,RV} (q)$ become relevant:
\bea
\tilde{g}_{LV,RV}^{(p)} &=& 2 \, g_{LV,RV}(u) + g_{LV,RV}(d)  \ , \\
\tilde{g}_{LV,RV}^{(n)} &=&  g_{LV,RV}(u) + 2\, g_{LV,RV}(d)   \ . 
\eea

\begin{table*}
\caption{\label{tab:table1} Vector and Axial-Vector couplings 
of $u$ and $d$ quarks to $Z_1$, $Z_2$, and $\gamma$. 
We list for completeness the effective 
Vector and Axial-Vector couplings induced by box diagrams of Fig. 
\ref{fig:boxes}.}
\begin{ruledtabular}
\begin{tabular}{cccccccc}
\multicolumn{2}{c}{$Z_1$}&\multicolumn{2}{c}{$Z_2$}&
\multicolumn{2}{c}{$\gamma$} &
\multicolumn{2}{c}{BOX}
\\
\hline
\multicolumn{2}{c}{$v_u^{(1)}  = 1 - \frac{8}{3} s^2_W   $} 
& 
\multicolumn{2}{c}{    
$v_u^{(2)} =1 - \frac{8}{3} s^2_W $}& 
\multicolumn{2}{c}{    
$v_u^{(\gamma)}=   \frac{2}{3} $} &
\multicolumn{2}{c}{    
$v_u^{(\rm B)}= 1 $} 
\\ 
\multicolumn{2}{c}{    
$a_u^{(1)}= 1   $} & 
\multicolumn{2}{c}{ $a_u^{(2)}= -1 + 2 s^2_W  $}
 &  \multicolumn{2}{c}{    $  $}  &
\multicolumn{2}{c}{    
$a_u^{(\rm B)}= - 1 $} 
\\ 
\hline
\multicolumn{2}{c}{    
$v_d^{(1)}= - 1 + \frac{4}{3}  s^2_W $}
& \multicolumn{2}{c}{
$v_d^{(2)} = -1 + \frac{4}{3}  s^2_W  $} & 
\multicolumn{2}{c}{    
$v_d^{(\gamma)}= -\frac{1}{3}   $} &
\multicolumn{2}{c}{    
$v_d^{(\rm B)}= -\frac{1}{4} $} 
\\ 
\multicolumn{2}{c}{    
$a_d^{(1)} = -1    $} & 
\multicolumn{2}{c}{ $a_d^{(2)}= 1 - 2 s^2_W $} 
&  \multicolumn{2}{c}{    $  $} 
&    
\multicolumn{2}{c}{    
$a_d^{(\rm B)}= \frac{1}{4} $} 
\\ 
\end{tabular}
\end{ruledtabular}
\end{table*}

\subsection{Effective Lagrangian for $\mu  \rightarrow 3 e$}
The process $\mu \rightarrow 3 e$ can occur in the LRSM through (i)
tree level exchange of doubly charged Higgses (via the interaction of
Eq.(\ref{eq:xlag2})); (ii) one-loop effective $\mu \rightarrow e$
vertex, with an electron line attached to the gauge boson; (iii) box
diagrams.  Barring the unnatural possibility that
$M_{\delta_{L,R}^{\pm \pm}} \gg M_{W_2}$, the loop amplitudes (ii) 
and (iii) are
suppressed by the standard $\alpha/\pi$ factor, and therefore in our
analysis we disregard them.

Doubly charged Higgs particles mediate at tree level also the 
decays $\tau \rightarrow l_a \, l_b \, \bar{l}_c$, with 
$l_{a,b,c}=\mu, e$. In compact notation, the effective 
lagrangian for four-lepton processes is given by:
\beq
{\cal L}_{\delta} = \frac{g^2}{4} h_{ij} h_{km}^*  \ 
\left[ \frac{1}{M^2_{\delta_R^{++}}}  
\left( \overline{l_{iR}^c} \, l_{j R} \right) \, 
\left( \overline{l_{k R}} \, l_{m R}^c \right)
+  ( L \leftrightarrow R) 
\right]\ . 
\eeq

\section{Analysis}  

Based on the results described in the previous section, we now discuss
the phenomenology of lepton flavor violation in muon decays within the
LRSM.  
There are three main objectives of our analysis. First, we shall
identify relations between LFV rates that are largely independent of
the model parameters, and therefore can be considered as signatures of
left-right symmetry broken at the multi-TeV scale. The pattern
emerging is remarkably clear, and could be confronted with
experimental findings in the next decade: the branching fractions for
$\mu \rightarrow e$ conversion and $\mu \rightarrow e \gamma$ are
expected to be very similar, and two order of magnitude smaller than
the one for $\mu \rightarrow 3 e$ (with some caveats).
Second, we shall study the constraints on heavy neutrino masses and
mixings implied by present experimental limits on LFV processes.
And third, we shall discuss 
the impact of future experiments, including collider measurements.

Before describing the details of our analysis let us shortly recall the
existing limits on the model parameters of interest to us.  Direct
searches imply that $M_{W_2} \geq 786$ GeV, while singly- and
doubly-charged Higgs particles should be heavier than $\sim$100
GeV~\cite{Hagiwara:fs}. Indirect bounds are stronger and require the
Higgs masses to be on the TeV scale. 
In summary, the existing phenomenology is consistent with the heavy
sector masses being generically at the TeV scale or above. 
In what follows, we shall 
explore the consequences of a heavy mass scale being in the range 
1-10 TeV,  which can be tested in the foreseeable future.

\subsection{Setting the stage}

The quantities of primary interest to us are the branching ratios:
\beq
B_{\mu \rightarrow e \gamma} = 
\frac{\Gamma (\mu \rightarrow e \gamma)}{ \Gamma_\mu^{(0)}}  \ , 
\qquad  
B_{\mu \to e}^{Z} = 
\frac{\Gamma_{\rm conv}^{Z}}{\Gamma_{\rm capt}^{Z}} \ , 
\qquad 
B_{\mu \rightarrow 3 e} = 
\frac{\Gamma (\mu \rightarrow 3 e)}{ \Gamma_\mu^{(0)}}  \ , 
\eeq
where $\Gamma_\mu^{(0)} = (G_F^2 m_\mu^5)/(192 \pi^3) $,  and for the
capture rate $\Gamma_{\rm capt}^{Z}$ we take the experimental values. 
The expression for the conversion rate $\Gamma_{\rm conv}^{Z}$ 
involves the overlap integrals~\cite{Kitano:2002mt} 
\bea
V^{(p,n)} &=& \displaystyle\frac{1}{2 \sqrt{2}} \,  
\int_{0}^{\infty} \ dr  \, r^2 \, N^{(p,n)} \rho^{(p,n)} \, 
\left( g_{e}^- g_{\mu}^- + f_{e}^- f_{\mu}^-  \right)   \ , 
\\
D &=& - \displaystyle\frac{4 m_\mu }{\sqrt{2}} \,  
\int_{0}^{\infty} \ dr \,  r^2 \, E(r) \, 
\left( g_{e}^- f_{\mu}^- + f_{e}^- g_{\mu}^-  \right)  \ . 
\eea
Here $N^{(p)} = Z$, $N^{(n)} = A-Z$; $\rho^{(p,n)}$ are proton and
neutron densities, $E(r)$ is the electric field generated by protons,
and $g^{-}_{\mu,e}$ $f^{-}_{\mu,e}$, are the upper and lower
components of the initial bound muon and final continuum 
electron wavefunctions, obtained by solving
the Dirac equation. The overlap integrals have dimension of (mass)$^{5/2}$, 
and in our study we use the numerical results for them
reported in Table I of Ref.~\cite{Kitano:2002mt}. 
In terms of the form factors calculated above and $D, V^{(n)},
V^{(p)}$, the relevant branching fractions read:
\bea
B_{\mu \rightarrow e \gamma} &=& 384 \pi^2  \, e^2  (|A_L|^2 + |A_R|^2)  
\ , \label{eq:muegBR} 
\\
B_{\mu \to e} &=& \displaystyle\frac{2 \, G_F^2}{\Gamma_{\rm capt}} \,   
\left(   | A_R^*  \, D + \tilde{g}_{LV}^{(p)} V^{(p)} + 
 \tilde{g}_{LV}^{(n)} V^{(n)} |^2   +  
| A_L^*  \, D + \tilde{g}_{RV}^{(p)} V^{(p)} + 
 \tilde{g}_{RV}^{(n)} V^{(n)} |^2   \right)   \ , 
\label{eq:muconvBR}
\\
B_{\mu \rightarrow 3 e} &=& \frac{1}{2} \, | h_{\mu e} h_{ee}^* |^2   
\left( \frac{M_{W_1}^4}{M_{\delta_{L}^{++}}^4} +
\frac{M_{W_1}^4}{M_{\delta_{R}^{++}}^4} \right) \ . 
\label{eq:mu3eBR}
\eea
While $B_{\mu \rightarrow 3 e}$ has a relatively simple structure,  
in general 
$B_{\mu \rightarrow e \gamma}$ and $B_{\mu \to e}$
depend on a large number of unknown model parameters. However, under
the rather natural assumption  
of a "commensurate mass spectrum" for the heavy sector of the model  
({\em i.e.}, $M_{W_2}\sim M_{\delta_R^{++}}
\sim M_{\delta_L^{++}} \sim M_{H^+}$), the problem becomes more tractable. 
Specifically, if $M_{W_2}$, $M_{\delta_R^{++}}$, 
$M_{\delta_L^{++}}$, $M_{H^{+}}$, and the heavy neutrino masses $M_n$ 
are all of the same order of magnitude (in practice we
shall assume $0.2 \lsim M_i/M_j \lsim 5$ for each pair of masses), 
the amplitudes for $\mu \rightarrow e$  and $\mu \rightarrow e \gamma$   
become approximately proportional to $g_{\rm lfv}$, defined in 
Eq.~(\ref{eq:xcouplings}).  
This is based on the following observations:
\begin{itemize} 
\item[(i)] Doubly charged  Higgs contributions to the couplings 
$A_{L,R}$, $g_{LV}(q)$ and $g_{RV}(q)$ are linear in $x_n$  
(hence proportional  to $g_{\rm lfv}$), and are sizable 
(the anapole transition form factor receives a large logarithmic enhancement). 
\item[(ii)] Gauge contributions depend on $x_n$ through the 
functions $S_i (x)$. These terms always represent a small correction 
to the Higgs contribution because 
$$ 
(a): \  \left| S_i (x) \right| \ll x \ 
\frac{8}{3} \log \frac{M_{\delta^{++}}^2}{m_\mu^2} ;   \qquad \qquad    
(b): \  \left| S_i^{'} (x) \right| \ll 
\frac{8}{3} \log \frac{M_{\delta^{++}}^2}{m_\mu^2} ;   \qquad \qquad    
$$
within the region $ (0.2)^2 \leq x \leq 3 $, 
where the lower limit follows from  our
assumption of commensurate spectrum and the upper limit from the vacuum 
stability condition~\cite{Mohapatra:pj,Prezeau:2003xn}
\footnote{In the case of $A_L$, the relevant conditions are $|S_3(x)|
\ll x/3$, and $|S_3^{'} (x)| \ll 1/3$.}.
\end{itemize}
Condition {\it (a)} ensures that gauge terms are small in the 
case of non-degenerate heavy neutrinos, while condition {\it (b)} 
suppresses them 
in the case of nearly degenerate neutrinos. 
In what follows we account for the small gauge-induced contributions 
to the various couplings  by expanding the $S_i(x)$ around
$\bar{x}=1.5$, and keeping only the linear term.   
We have checked that  
the residual dependence on the expansion point $\bar{x}$ is small, 
and does not affect  our discussion and results in a significant way. 

The above considerations about the relevance of $g_{\rm lfv}$ remain
true even in the unnatural limit $M_{\delta_{L,R}^{++}} \ll M_{W_2}$, but
become invalid in the opposite limit $M_{\delta_{L,R}^{++}} \gg M_{W_2}$, as
for $M_{\delta_{L,R}^{++}} \sim 10 M_{W_2}$ the Higgs mass suppression 
compensates the logarithmic enhancement.  
Such unnatural limit will not be considered here.  

In summary, in the natural scenario of commensurate mass spectrum
in the heavy sector, $B_{\mu \to e}^Z$ and $B_{\mu \rightarrow e \gamma}$
are driven by a single combination of heavy neutrino masses and 
mixing parameters, which we defined as $g_{\rm lfv}$. Moreover, 
$B_{\mu \to e}^Z$ and $B_{\mu \rightarrow e \gamma}$ 
depend only on four independent parameters ($g_{\rm lfv}, M_{W_2},
M_{\delta_L^{++}},M_{\delta_R^{++}}$), and have the generic structure
\beq
B_i = |g_{\rm lfv}|^2 \ \frac{M_{W_1}^4}{M_{W_2}^4} \times  
f_{i} \left( 
\log \frac{M_{W_2}}{\sqrt{- q^2}} , 
\log \frac{M_{W_2}}{M_{W_1}} , 
r_L \equiv  \frac{M_{\delta_L^{++}}}{M_{W_2}}, 
r_R \equiv  \frac{M_{\delta_R^{++}}}{M_{W_2}} \right) \ .  
\label{eq:pheno1}
\eeq 
We shall next explore the consequences of such simplified form. 

\subsection{$\mu \rightarrow e$ conversion versus $\mu \rightarrow e \gamma$} 

The first important consequence is that the ratio $R^A \equiv
B_{\mu \to e}^A/B_{\mu \rightarrow e \gamma}$ does not depend on $g_{\rm
lfv}$, and is a function of $\log (M_{W_2}/M_{W_1}), r_L, r_R$.  
\begin{figure}[!b]
\begin{center}
\leavevmode
\begin{picture}(180,200)
\put(40,60){\makebox(70,70){\includegraphics[width=10.0cm]{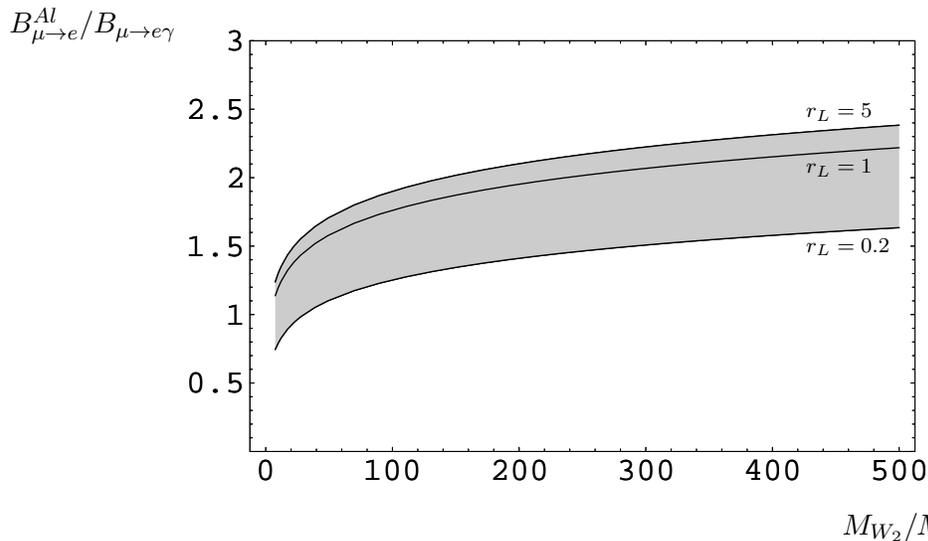}}}
\put(-135,180){  $B_{\mu \to e}^{Al}/B_{\mu \rightarrow e \gamma}$  }
\put(180,-10){ $M_{W_2}/M_{W_1}$ }
%
%
\put(170,97){{\scriptsize $r_{L}=0.2$} }
\put(170,127){{\scriptsize $r_{L}=1$} }
\put(170,148){{\scriptsize $r_{L}=5$} }
\end{picture}
\caption{  
$R^{Al} \equiv B_{\mu \to e}^{Al}/B_{\mu \rightarrow e \gamma}$ as a
function of $M_{W_2}/M_{W_1}$, for different values of $r_{L}$. We
keep $r_R=1$, because the variation of $R^{Al}$ with this parameter is
considerably smaller then the variation with $r_L$.  The shaded band can be
considered a prediction of left-right symmetry for $R^{Al}$, assuming 
commensurate heavy sector.}
\label{fig:RAl}
\end{center}
\end{figure}  
Our explicit analysis shows that, for input parameters in the
commensurate range $0.2 \leq r_{L,R} \leq 5$, $R^A$ varies at most
by $30 \%$ for any fixed value of $M_{W_2}/M_{W_1}$.
As illustration, in Fig.~\ref{fig:RAl} we show the ratio $R$ for 
aluminum $R^{Al}$ 
as a function of $M_{W_2}/M_{W_1}$ for a range of $r_{L}$ values 
(the variation with $r_R$ is much smaller).  The most striking feature
of our result is the near independence on the heavy mass parameters (as
long as they stay in the natural range), leading to a distinctive
prediction of the LRSM for $R^{Al}$. This ratio is of
${\cal O}(1)$ in this model and it is naturally confined between 1 and
2, as shown by the gray area in Fig.\ref{fig:RAl}.  The absolute scale
on this plot can be understood as a consequence of the logarithmic
enhancement of the anapole form factor contributing to $B_{\mu \to e}$.
Different values of $R^A$ (in particular values smaller than unity)
can be hardly accommodated without unnatural tuning of mass parameters.
Indeed, for mass parameters just above the present direct limits
($M_{W_2}=0.8$ TeV and $M_{\delta_{L,R}^{++}}=200$ GeV), we find
$R^{Al}=0.8$, which can be considered the minimal acceptable value
within this model.  This prediction is substantially different from 
R-parity conserving SUSY scenarios, and
can be hopefully tested by future measurements of $B_{\mu \to e}^{Al}$ 
(MECO) and $B_{\mu \rightarrow e \gamma}$ (MEG).

The qualitative features encountered in the analysis of $R^{Al}$ 
apply to other elements as well. In particular, the ratio $R^A$ is 
always of ${\cal O}(1)$. We have studied a few more examples, 
in the same range of mass parameters used above, finding: 
\beq
R^{Ti}:  2 \rightarrow 3.5 \ , \qquad
R^{Au}: 2 \rightarrow 4  \ , \qquad
R^{Pb}: 1.5 \rightarrow  3   \ . 
\eeq

\subsection{$\mu \rightarrow e$ conversion versus $\mu \rightarrow 3 e$}  

Under slightly stronger assumptions, it is also possible to derive an
order-of-magnitude relation between $B_{\mu \rightarrow 3 e}$ and
$B_{\mu \to e}$. Assuming dominance of logarithmic terms induced by doubly
charged Higgs diagrams, and using $ \log
(M^2_{\delta_{R}^{++}}/m_\mu^2) \approx \log
(M^2_{\delta_{L}^{++}}/{m_\mu^2})$, one can write
\beq
B_{\mu \to e} = \frac{8 G_F^2  \alpha^2}{9 \pi^2} \, 
\frac{(V^{(p)})^2}{\Gamma_{\rm capt}} \, 
\left( \frac{M_{W_1}^4}{M_{\delta_{L}^{++}}^4} +
\frac{M_{W_1}^4}{M_{\delta_{R}^{++}}^4} \right) 
\left(\log \frac{M^2_{\delta_{R}^{++}}}{m_\mu^2} \right)^2 \ 
\Bigg| 
h_{\mu e} h_{ee}^* + h_{\mu \mu} h_{\mu e}^*  + h_{\mu \tau } h_{\tau e}^* 
\Bigg|^2    \ . 
\eeq
Under the assumption that $h_{\mu e} h_{ee}^* \sim
h_{\mu \mu} h_{\mu e}^* \sim h_{\mu \tau } h_{\tau e}^* $, and that 
no cancellations occur between the three contributions, 
one then expects
\beq
B_{\mu \to e} =
k_f \, 
\frac{16 G_F^2  \alpha^2}{9 \pi^2} \, 
\frac{(V^{(p)})^2}{\Gamma_{\rm capt}} \, 
\left(\log \frac{M^2_{\delta_{R}^{++}}}{m_\mu^2} \right)^2 \ 
B_{\mu \rightarrow 3 e} \ ,  
\eeq    
where $k_f =|g_{{\rm lfv}}|^2/|h_{\mu e} h^*_{ee}|^2$ 
is a number of order 1. 
For $M^2_{\delta_{R}^{++}} \approx M^2_{\delta_{L}^{++}} \approx 1
{\rm TeV}$, this translates into 
\beq 
B_{\mu  \rightarrow 3 e}   \sim 
\frac{ 3 \times 10^{2}}{ k_f }  \ 
B_{\mu \to e}^{Al} \ . 
\eeq
So, within this model, one expects that $\mu \rightarrow 3 e$  
could be the first rare muon decay to be observed. 
Sizable deviations from the above pattern would provide information 
about the parameters $h_{\mu l} h_{l e}^*$.  In particular,
$ B_{\mu \to e}^{Al}/B_{\mu \rightarrow 3 e} \gg \ 10^{-2} $ would imply
dominance of the $l=\mu$ and/or $l=\tau$ contribution in $|\sum_{l} h_{\mu
l} h_{l e}^*|$, and may lead to observable signals in $\tau \rightarrow
l_a l_b \bar{l}_c$ decays. 
On the other hand, $ B_{\mu \to e}^{Al}/B_{\mu \rightarrow 3 e} \ll \
10^{-3} $ would signal non-trivial relative phases among the
couplings, necessary to suppress $|\sum_{l} h_{\mu l} h_{l e}^*|$
compared to $|h_{\mu e} h_{e e}^*|$.

\subsection{Constraints on heavy neutrino masses and mixing}

\begin{figure}[!b]
\begin{center}
\leavevmode
\begin{picture}(150,180)
\put(-80,50){\makebox(70,70){\includegraphics[width=8.0cm]{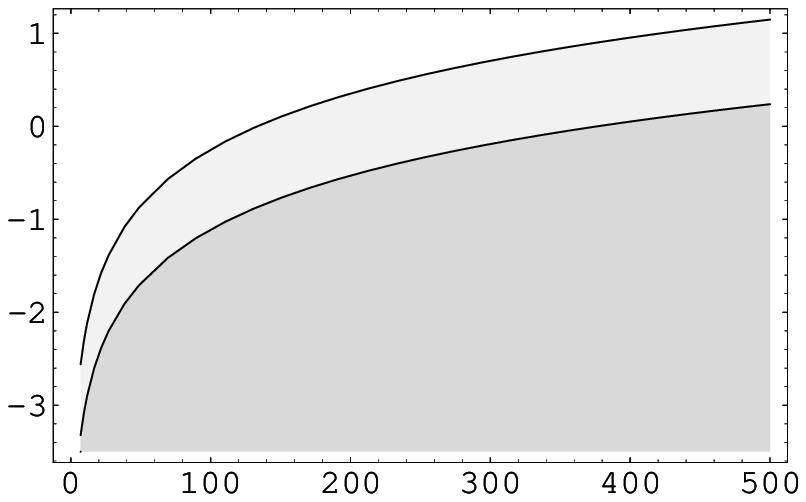}}}
\put(170,50){\makebox(70,70){\includegraphics[width=8.0cm]{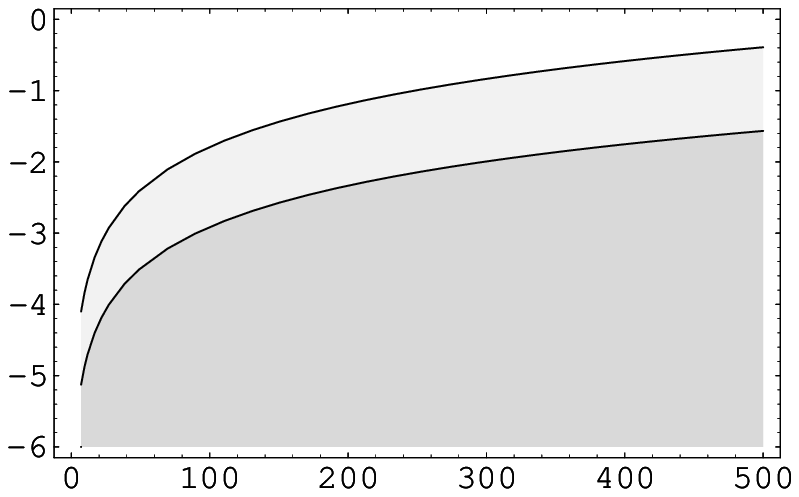}}}
\put(-170,160){  $ \log_{10} (g_{\rm lfv})$  }
\put(80,160){  $ \log_{10} (g_{\rm lfv})$  }
\put(-90,160){
\framebox[1.0\width][c]{
\small PRESENT LIMITS 
}}
\put(160,160){
\framebox[1.0\width][c]{
\small MECO AND MEG
}}
\put(-10,40){
\framebox[1.0\width][c]{
\scriptsize $r_L = r_R =1$
}}
\put(240,40){
\framebox[1.0\width][c]{
\scriptsize $r_L = r_R =1$
}}
\put(25,0){$M_{W_2}/M_{W_1}$}
\put(270,0){$M_{W_2}/M_{W_1}$}
\put(195,120){\scriptsize $B_{\mu \rightarrow e \gamma} < 10^{-14}$}  
\put(155,85){\scriptsize $B_{\mu \to e}^{\rm Al} < 10^{-16}$}
\put(-40,130){\scriptsize $B_{\mu \rightarrow e \gamma} < 1.2 \times 10^{-11}$}
\put(-50,100){\scriptsize $B_{\mu \to e}^{\rm Au} < 8 \times 10^{-13}$}
\end{picture}
\caption{
Correlations in the  $g_{\rm lfv}$-$(M_{W_2}/M_{W_1})$ plane 
imposed by present and future (MEG and MECO) limits on 
$B_{\mu \to e}$ and $B_{\mu \rightarrow e \gamma}$. 
The shaded area represents the region allowed by limits reported on the plot. 
In this plot we use $r_L=r_R=1$. 
Lowering $r_L$ and/or $r_R$ poses tighter constraints on $g_{\rm lfv}$, 
for fixed $M_{W_2}$. }
\label{fig:comp1}
\end{center}
\end{figure}

LFV in muon decays is driven by $g_{\rm lfv}$ and the couplings $h_{i
j}$, related to heavy neutrino masses and mixing angles through
Eqs.~(\ref{eq:hcouplings1})-(\ref{eq:xcouplings}).  We now explore the
correlations between $g_{\rm lfv}$ and heavy mass parameters implied
by present experimental limits and future limits/observations of
$B_{\mu \to e}^Z$, $B_{\mu \rightarrow e \gamma}$.  Subsequently, we
discuss the constraints on $h_{\mu e} h_{ee}^*$ implied by 
limits on $B_{\mu \rightarrow 3 e}$.

In order to illustrate the generic model expectations for 
$\mu \rightarrow e$ conversion and $\mu \rightarrow e \gamma$, we 
show below  approximate expressions for the rates (obtained by  
setting  $r_L=r_R=1$), 
\bea
B_{\mu  \rightarrow e \gamma} &=&  
1.5 \times 10^{-7}  
\  |g_{\rm lfv}|^2  \ 
\left(\frac{1 {\rm TeV}}{M_{W_2}}\right)^4  \ , 
\\
B_{\mu \to e}^{A,Z} &=&  
 X_A \times 10^{-7} 
\  |g_{\rm lfv}|^2  \ 
\left(\frac{1 {\rm TeV}}{M_{\delta^{++}_{L,R }}}\right)^4 \, 
\alpha \,
\left( \log \frac{M_{\delta^{++}_{L,R }}^2
}{m_\mu^2}   \right)^2 
\ , 
\eea
where $X_A$ is the nucleus dependent numerical factor  
(we find  $X_A$ = 0.8, 1.3, 1.6, and 1.1 for Al, Ti, Au, and Pb,
respectively). These branching ratios
have to be compared with present experimental limits:
\beq
B_{\mu \rightarrow e \gamma} < 1.2 \times  10^{-11} \,  
\mbox{\cite{muegamma99}}, 
\ 
B_{\mu \to e}^{Ti} < 4.3 \times 10^{-12}  \,  \mbox{\cite{mueconvTi}}, 
\ 
B_{\mu \to e}^{Au} <  8 \times 10^{-13} ~ \mbox{\cite{mueconvAu}},    
\ 
B_{\mu \to e}^{Pb} <    4.6 \times 10^{-11} ~ \mbox{\cite{mueconvPb}}     \ . 
\eeq 
Thus, assuming commensurate spectrum and $g_{\rm lfv} \sim 1$
(i.e. large mixing angles and non-degenerate heavy neutrinos),
consistency with present limits implies that the scale of $SU(2)_R$
breaking has to be around 20 TeV.  On the other hand, for $M_{W_2}$ in
the 1-10 TeV range,  present experimental limits already  
impose non-trivial constraints on $g_{\rm lfv}$ (left panel in
Fig.~\ref{fig:comp1}).  Values of $g_{\rm lfv}$ at the $10^{-2} -
10^{-3}$ level imply either small mixing angles in the heavy neutrino
sector or nearly degenerate heavy neutrinos, on the scale set by
$M_{W_2}$.  The most stringent constraints at present come from $\mu
\rightarrow e$ conversion in gold.  Future experiments MEG~\cite{MEG}
and MECO~\cite{MECO} will be able to probe even higher mass scales and
put more stringent upper limits on $g_{\rm lfv}$ (right panel in
Fig. \ref{fig:comp1}).  Once again, $\mu \rightarrow e$ conversion 
will probe the model parameter space more strongly.

\begin{figure}[!b]
\begin{center}
\leavevmode
\begin{picture}(150,180)
\put(-80,50){\makebox(70,70){\includegraphics[width=8.0cm]{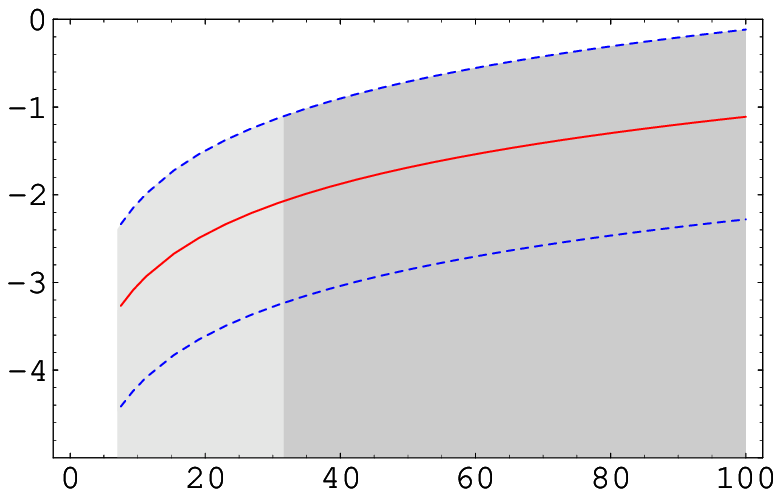}}}
\put(170,50){\makebox(70,70){\includegraphics[width=8.0cm]{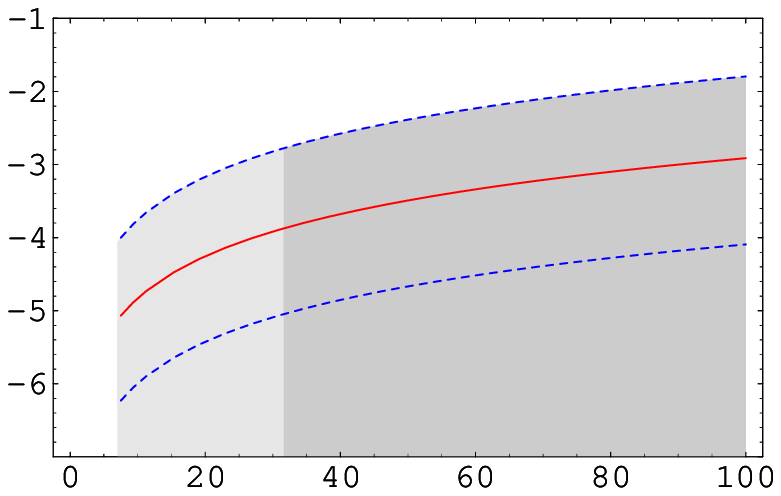}}}
\put(-170,160){  $ \log_{10} (g_{\rm lfv})$  }
\put(80,160){  $ \log_{10} (g_{\rm lfv})$  }
\put(-80,160){
\framebox[1.0\width][c]{
$B_{\mu \to e}^{Au} <  8 \times 10^{-13}$}}
\put(170,160){
\framebox[1.0\width][c]{
$B_{\mu \to e}^{Al} < 10^{-16}$}}
\put(25,0){$M_{W_2}/M_{W_1}$}
\put(270,0){$M_{W_2}/M_{W_1}$}
\put(-115,35){{\scriptsize $\xi = 10^{-3}$}}
\put(135,35){{\scriptsize $\xi = 10^{-3}$}}
\put(270,90){{\scriptsize $r_{L.R}=0.2$} }
\put(270,115){{\scriptsize $r_{L,R}=1$} }
\put(270,135){{\scriptsize $r_{L,R}=5$} }
\put(10,80){{\scriptsize $r_{L,R}=0.2$} }
\put(10,110){{\scriptsize $r_{L,R}=1$} }
\put(10,135){{\scriptsize $r_{L,R}=5$} }
\end{picture}
\caption{
Correlations in the $g_{\rm lfv}$-$(M_{W_2}/M_{W_1})$ plane imposed by
$B_{\mu \to e}$, before and after MECO's goal has been reached, 
for different values of
$r_{L}=r_{R}$.  The shaded area represents the region allowed by the
assumed limits on $B_{\mu \to e}$.  A non-zero mixing angle $\xi$ would 
further reduce the allowed region.  As an illustration, 
the allowed region for $\xi = 10^{-3}$ is plotted in light-gray.}
\label{fig:comp2}
\end{center}
\end{figure} 

Focusing on $\mu \rightarrow e$ conversion (present limits and
projected MECO sensitivity), in Fig.~\ref{fig:comp2} we report a more
detailed study of the constraints.  At fixed $M_{W_2}$, lowering or
raising $r_{L,R}$ within the natural range $0.2 \lsim r_{L,R} \lsim 5$,
can change the bound on $g_{\rm lfv}$ by an order of magnitude.
Lighter Higgs particles imply tighter upper limits on $g_{\rm lfv}$.
Finally, the impact of a non-zero mixing angle $\xi$ (detectable, 
for example, through right-handed current signals in $\beta$ decays) is also
considered in Fig.\ref{fig:comp2}. A non-vanishing $\xi$ would
imply~\cite{Masso:1984bt} the upper bound $M_{W_2}/M_{W_1} \leq
1/\sqrt{\xi}$, and thus narrow down the allowed region in the $g_{\rm
lfv}$-$M_{W_2}/M_{W_1}$ plane (light-gray region in
Fig.\ref{fig:comp2}).

Additional information on heavy neutrino parameters can be obtained in
principle from $\mu \rightarrow 3 e$.  The rate depends on doubly
charged Higgs masses and the combination $|h_{\mu e} \, h_{ee}^* |$
(Eq.~(\ref{eq:mu3eBR})).  The present limit $B_{\mu \rightarrow 3 e} <
10^{-12}$ \cite{Hagiwara:fs} implies (assuming $M_{\delta_L^{++}} =
M_{\delta_R^{++}} $) \footnote{ A weaker upper limit on the same
combination of parameters can be derived from searches of muonium
anti-muonium transition~\cite{Abela:dm,Herczeg:1992pt}.  
In general, present limits on the flavor
diagonal coupling $h_{ee}$ from Bhabha scattering~\cite{Kuze:2002vb}, and other
combinations of $h_{ij}$ from rare $\tau$ decays are much weaker
(typically $B_{\tau \rightarrow l_a l_b l_c} < 10^{-6}$~\cite{Hagiwara:fs}).}
\beq 
|h_{\mu e} \, h_{ee}^* | \leq 1.55 \times
10^{-4} \, \sqrt{ \frac{B_{\mu \rightarrow 3 e}}{10^{-12}}} \, \left(
\frac{M_{\delta_{L,R}^{++}}}{1 {\rm TeV}} \right)^2 \ .
\label{eq:hcouplings2}
\eeq
Thus, assuming $M_{\delta^{++}} \sim 1 \, {\rm TeV} $, the couplings
$h_{i j}$ are constrained to be at the $\sim 10^{-2}$ level. 
Unlike the case of $g_{\rm lfv}$, however, the smallness of 
$h_{\mu e}$ does not imply small mixing angles or almost-degenerate  
heavy neutrinos, because the Majorana phases contained in $K_R$   
may lead to cancellations in the sum of Eq.~(\ref{eq:hcouplings1}).

\subsection{Testing the model: interplay with collider measurements}

As noted above, information from LFV processes and other aspects of
low energy phenomenology (such as signals of right-handed currents)
can severely constrain the model parameter space in the near future.
Moreover, given that $B_{\mu \to e}$ and $B_{\mu \rightarrow e \gamma}$
depend only on $g_{\rm lfv}, M_{W_2}, M_{\delta_L^{++}},
M_{\delta_R^{++}}$, collider searches of heavy particles and low 
energy searches of LFV decays jointly provide a powerful probe of
left-right symmetry.  In fact, in the best-case scenario, separate
measurements of $B_{\mu \to e}^{Al}$, $B_{\mu \rightarrow e \gamma}$ and
the mass parameters $M_{W_2}, M_{\delta_L^{++}},M_{\delta_R^{++}}$
would allow one to test the model (four parameters versus five
observables).  Even in less optimistic scenarios, one can imagine
using collider information to narrow down the model predictions for
LFV processes, or use observation of LFV to determine 
allowed regions in the heavy mass parameter space. 
\begin{figure}[!t]
\begin{center}
\leavevmode
\begin{picture}(200,220)
\put(-60,70){\makebox(70,70){\includegraphics[width=6.5cm]{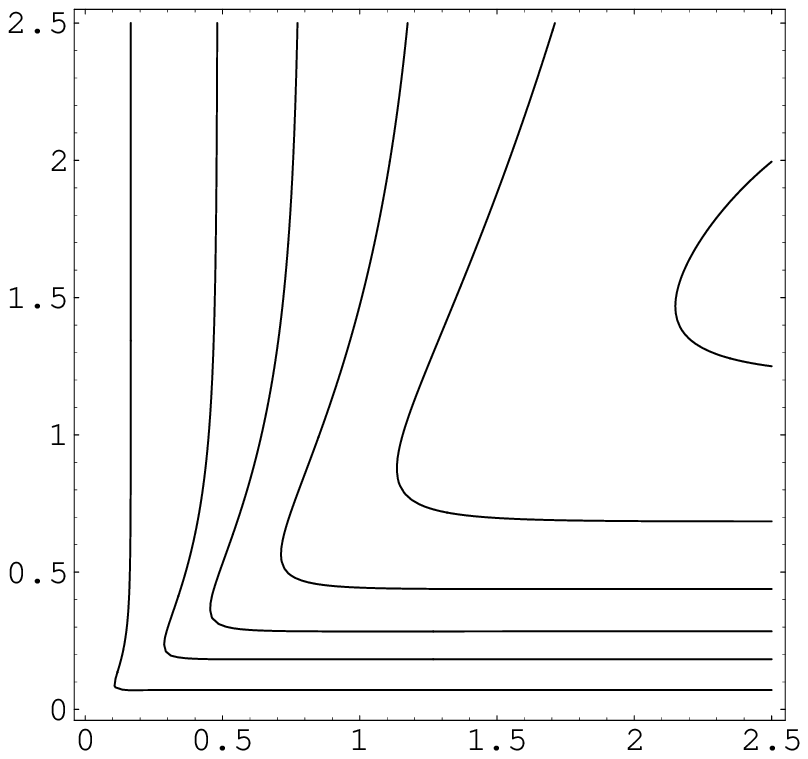}}}
\put(-130,200){  $M_{\delta_R^{++}} ({\rm TeV})$  }
\put(20,0){  $M_{\delta_L^{++}} ({\rm TeV})$  }
\put(180,70){\makebox(70,70){\includegraphics[width=6.5cm]{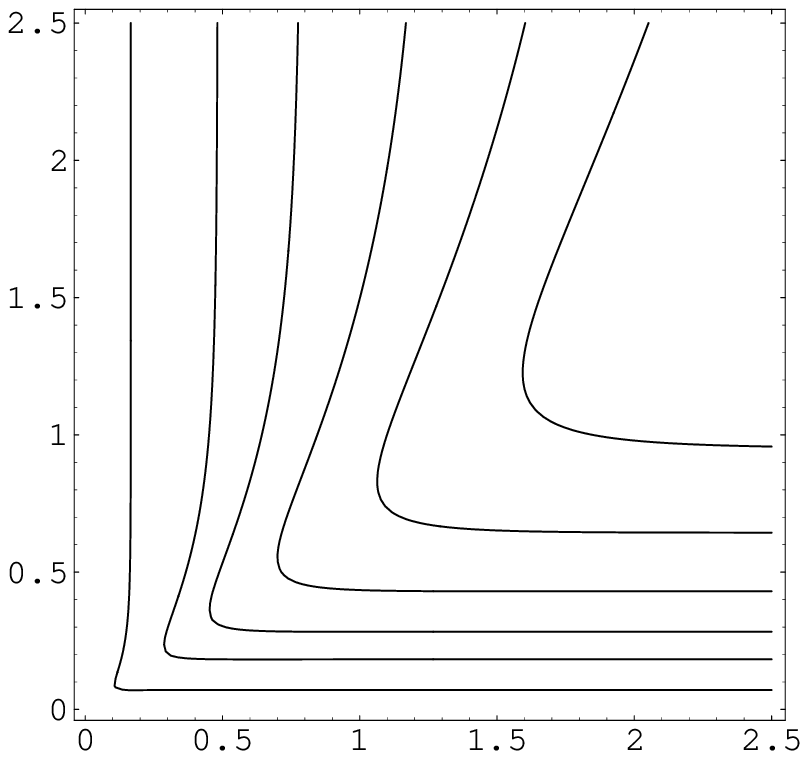}}}
\put(105,200){  $M_{\delta_R^{++}} ({\rm TeV})$  }
\put(260,0){  $M_{\delta_L^{++}} ({\rm TeV})$  }
\put(-40,200){
\framebox[1.0\width][c]{
\scriptsize 
$M_{W_2} = 1\, {\rm TeV}$
}
}
\put(195,200){
\framebox[1.0\width][c]{
\scriptsize 
$M_{W_2} = 5 \, {\rm TeV}$
}
}
\put(-88,170){{\scriptsize $0.8$}}
\put(-67,170){{\scriptsize $1.0$}}
\put(-48,170){{\scriptsize $1.1$}}
\put(-24,170){{\scriptsize $1.2$}}
\put(7,170){{\scriptsize $1.3$}}
\put(37,150){{\scriptsize $1.4$}}
\put(153,170){{\scriptsize $0.8$}}
\put(173,170){{\scriptsize $1.0$}}
\put(192,170){{\scriptsize $1.1$}}
\put(215,170){{\scriptsize $1.2$}}
\put(240,170){{\scriptsize $1.3$}}
\put(268,170){{\scriptsize $1.4$}}
\end{picture}
\caption{
Contour plot of 
$R^{Al} \equiv B_{\mu \to e}^{\rm Al}/B_{\mu \rightarrow e \gamma} $
in the  $M_{\delta_L^{++}}$ -$M_{\delta_R^{++}}$ plane, 
for $M_{W_2}=1$ TeV (left panel) and 
for $M_{W_2}=5$ TeV (right panel). Each curve is labeled by the 
corresponding  $R^{Al}$.  
As a function of the Higgs mass along the line $M_{\delta_L^{++}} =
M_{\delta_R^{++}}$, $R^{Al}$ reaches a maximum at 
$M_{\delta_L^{++}} \sim 2 \,  M_{W_2}$ and then decreases, due to 
decoupling of doubly charged Higgs bosons (the latter effect is 
not visible in the plots).}
\label{fig:ill1}
\end{center}
\end{figure} 

As a simple illustration of this point, we show in
Fig.~\ref{fig:ill1} contour plots of $R^{Al} \equiv
B_{\mu \to e}^{Al}/B_{\mu \rightarrow e \gamma}$ in the
$M_{\delta^{++}_{L}}$-$M_{\delta^{++}_{R}}$ plane, for two values of
$M_{W_2}$. We focus on the case of heavy masses in the 1-2
TeV range,  which will be accessible at the LHC and 
Tevatron II~\cite{Datta:1999nc}. 
In this mass-region, the model expectations  
are almost independent of $M_{W_2}$. Moreover, one sees that values of
$R^{Al} < 0.8$ can only occur for $M_{\delta_{L,R}^{++}} < 100 \, {\rm
GeV}$, already excluded by direct searches.  
Depending on future experimental developments, possible uses of the
plots in Figs.~\ref{fig:ill1} include:
\begin{itemize}
\item Given measurements of Higgs and heavy gauge boson masses, one can  
infer rather precisely where to expect $R^{Al}$ within 
this scenario.
\item Given an experimental signal for $B_{\mu \to e}^{Al}$ and $B_{\mu
\rightarrow e \gamma}$, one can identify the allowed region in the
$M_{\delta^{++}_{L}}$-$M_{\delta^{++}_{R}}$ plane, for different
values of $M_{W_2}$.  Collider searches could then confirm or falsify
the model expectations. 
As can be seen from the plots, however, in order to have a significant test,  
the fractional uncertainty on $R^{Al}$ should be at most $20 \%$
(otherwise most of the $M_{\delta^{++}_{L}}$-$M_{\delta^{++}_{R}}$
would be allowed). 
Given the projected sensitivities, this may be achieved at the next
generation experiments if $B_{\mu \to e \gamma} \geq 2. \times
10^{-13}$. 

\end{itemize}

\section{Conclusions}
The study of flavor violation among leptons now lies at the forefront
of particle and nuclear physics. The tiny masses of the three lightest
neutrinos and the nearly maximal mixing among them stands in stark
contrast with the situation involving quarks, and the origin of this
difference remains a fundamental and unsolved puzzle. A variety of
scenarios have been proposed that attempt to answer this question, and
these ideas would have predictable consequences for other
observables. In this study, we have analyzed the consequences of one
such scenario -- the left-right symmetric model -- that entails
a minimal extension of the SM gauge symmetries and that includes
non-sterile, right-handed neutrinos whose mass could be generated at
the multi-TeV scale, albeit with some fine-tuning.  
We have shown how it implies relationships among
various LFV decays of the muon that could distinguish it
experimentally from other models of LFV. We have also illustrated how
direct searches for right-handed gauge bosons and triplet Higgs at the
Tevatron and LHC would complement the charged lepton LFV studies and
either help favor or rule out the possibility of rather low-scale LFV
without SUSY.

The main conclusions of our study are:
\begin{itemize}
\item The branching ratios $B_{\mu \rightarrow e}$ and 
$B_{\mu \rightarrow e \gamma}$ are similar in magnitude, in distinction
to other possible scenarios which predict that 
 $B_{\mu \rightarrow e}/B_{\mu \rightarrow e \gamma} \sim \alpha$.
\item Within the LRSM, and with reasonable additional assumptions, 
$B_{\mu \rightarrow 3e}/B_{\mu \rightarrow e} \sim$ 300, making
the process $\mu \rightarrow 3e$ perhaps easiest to observe.
\item The existing limits on the LFV muon decays already
substantially constrain the mixing and mass splittings of
the heavy right-handed neutrinos. The planned more sensitive
experiments will therefore test the LRSM severely.
\end{itemize}

If the LRSM scenario turns out to be correct, the deeper connections
between the heavy and light neutrino spectrum would, then, have to be
pursued by additional experimental and theoretical work. On the other
hand, should experiment eliminate the possibility of
non-supersymmetric, low-scale LFV based on the considerations
discussed above, the lepton flavor problem will nevertheless remain a
rich area of study, both theoretically and experimentally, for some
time to come.

\begin{acknowledgments}
We thank M.B Wise for useful comments provided during the course of
carrying out this calculation. This work was supported in part under
U.S. Department of Energy contract \# DE-FG03-88ER40397 and NSF Award
PHY-0071856.  V.C. was supported by a Sherman Fairchild Fellowship from 
Caltech. 
\end{acknowledgments}

\appendix

\section{Loop functions}  
\label{app:functions}    

We collect here the functions $S_{i}(x)$ and $D_{1}(x,y)$ 
appearing in the expression of various $\mu \rightarrow e$ form factors. 
\bea
S_1 (x) &=&  \frac{ 4 \, x}{(1 - x)^2} \, 
\left[ 
6 - 7 x + x^2 + (2 + 3 x) \log x 
\right]  \ , 
\\
S_2 (x) &=& \frac{x ( 4 - 3 x)}{(1 - x)^2}  - 
\frac{2 x ( 12 - 10 x + x^2)}{3 ( 1 - x)^2} \, ( S_4 (x) + 1 )  \ , 
\\   
S_3 (x) &=& 
- \frac{x ( 1 + 2 x)}{8 (1 - x)^2}  +
\frac{3 x^2}{4 (1 - x)^2} \, ( S_4 (x) + 1 )  \ , 
\\
S_4 (x) &=& \frac{x}{(1 - x)^2} \, (1 - x + \log x)  \ ,   
\\ 
D_{1} (x,y) &=&  x \left(2 - \log \frac{y}{x} \right) + 
\displaystyle\frac{ (-8 x + 9 x^2 - x^3) + (-8 x^2 + x^3) \log x}{ 
(1 - x)^2} + \displaystyle\frac{x ( y - y^2 + y^2 \log y)}{(1-y)^2} 
\nonumber \\
& & + \displaystyle\frac{2 x y (4 - x) \log x }{(1-x) (1-y)} + 
 \displaystyle\frac{2 x (x - 4 y) \log  (y/x)}{(x-y) (1-y)}  \  . 
\eea
Both $S_i(x)$ and $D_1(x,y)$ are regular at $x = 1$ and $y = 1$.
Note that the potentially dangerous contribution involving the 
large mass-ratio $y_n = (M_n/M_{W_1})^2$  has a finite limit 
for $y_n \rightarrow \infty$:
$$ \lim_{y \rightarrow \infty} D_{1}(x,y) = - 7 S_4 (x)  \ . $$

\section{Full expressions for $\mu \, e \, \gamma^* $  form factors} 
\label{app:mueg}

In terms of the interactions vertices reported in Eq.~(\ref{eq:xlag1},
\ref{eq:xlag2}), and without neglecting the charged lepton
mass-dependence of loops, the photonic form factors read:
\bea
F_R^{\rm (\gamma)} & = & 
\displaystyle\sum_{n={\rm heavy}} \, \Big( K_R^\dagger \Big)_{e n} 
\Big( K_R \Big)_{n \mu}  \ 
 \frac{M_{W_1}^2}{M_{W_2}^2}  \, S_2 (x_n) 
\nonumber \\
& &  + \displaystyle\sum_{l=e,\mu,\tau} \, 
h_{e l}^* \,  h_{l \mu} \ 
\frac{M_{W_1}^2}{M_{\delta_R^{++}}^2}  
\left[- \frac{40}{9}  -  \frac{8}{3} \log 
\left( \frac{-q^2}{M_{\delta_R^{++}}^2} \right)
- 16 \, S_5 \left( \frac{m_l^2}{-q^2} \right)
\right]  \ , 
\\
F_L^{\rm (\gamma)} & = & 
\displaystyle\sum_{l=e,\mu,\tau} \, 
h^{*}_{e l} \, h_{l \mu}
\ 
\frac{M_{W_1}^2}{M_{\delta_L^{++}}^2}  
\left[- \frac{40}{9}  -  \frac{8}{3} \log 
\left( \frac{-q^2}{M_{\delta_L^{++}}^2} \right)
- 16 \, S_5 \left( \frac{m_l^2}{-q^2} \right)
 \right] \nonumber \\ 
 & & - \frac{2}{9} 
\Big( \tilde{h}^\dagger \tilde{h} \Big)_{e \mu} 
\frac{M_{W_1}^2}{M_{\delta_L^{+}}^2}    \ , 
\\
16 \pi^2 \,  A_L & = & 
\displaystyle\sum_{n={\rm heavy}} \, \Big( K_R^\dagger \Big)_{e n} 
\Big( K_R \Big)_{n \mu}  \ 
 \frac{M_{W_1}^2}{M_{W_2}^2}  \, S_3 (x_n) 
 - \frac{1}{3}  \displaystyle\sum_{l=e,\mu,\tau} \ 
h^{*}_{e l} \, h_{l \mu}
\  \frac{M_{W_1}^2}{M_{\delta_R^{++}}^2}   \ , 
\\
16 \pi^2 \,  A_R & = & 
 - \frac{1}{3}  \displaystyle\sum_{l=e,\mu,\tau} \ 
 h^{*}_{e l}  \, h_{l \mu}
\  \frac{M_{W_1}^2}{M_{\delta_L^{++}}^2}  
- \frac{1}{24} 
 \Big( \tilde{h}^\dagger \tilde{h} \Big)_{e \mu}
\frac{M_{W_1}^2}{M_{\delta_L^{+}}^2} \ ,    
\eea
where the function $S_5(x)$ is:
\beq
S_5 (x) = \int_{0}^{1} d y \, y (1 - y) \, \log 
\Big( x + y (1-y) \Big)  \ . 
\eeq


\end{document}